%% file: paper.tex
\documentclass[sigconf]{acmart}

\usepackage{graphicx}
\usepackage{graphics}
\usepackage{mdwlist}
\usepackage{algorithm}
\usepackage[noend]{algpseudocode}
\usepackage{booktabs}
\usepackage[center]{subfigure}

\usepackage{amsthm,amssymb}
\usepackage[]{caption}
\usepackage{multirow}
\usepackage{mathrsfs}
\usepackage{amsbsy}
\usepackage[mathscr]{eucal} 
\usepackage{xspace}
\usepackage{paralist} 

\usepackage{verbatimbox}
\usepackage{pifont}
\usepackage{balance}

\input{dfn.tex}

\begin{document}

\copyrightyear{2017} 
\acmYear{2017} 
\setcopyright{acmcopyright}
\acmConference{KDD'17}{}{August 13--17, 2017, Halifax, NS, Canada.}
\acmPrice{15.00}
\acmDOI{10.1145/3097983.3098087}
\acmISBN{978-1-4503-4887-4/17/08}

\fancyhead{}
\settopmatter{printacmref=false, printfolios=false}

\title{DenseAlert: Incremental Dense-Subtensor Detection in Tensor Streams}

\author{Kijung Shin, Bryan Hooi, Jisu Kim, Christos Faloutsos}
\affiliation{%
	\institution{School of Computer Science, Carnegie Mellon University, Pittsburgh, PA, USA}
}
\email{{kijungs,christos}@cs.cmu.edu,{bhooi,jisuk1}@andrew.cmu.edu}

\begin{abstract}
	\input{000abstract.tex}
\end{abstract}

\maketitle



\section{Introduction}
\label{sec:intro}
\input{010intro}

\section{Notations and Definitions}
\label{sec:prelim}
\input{020prelim}

\section{Proposed Method}
\label{sec:method}
\input{030method}

\section{Experiments}
\label{sec:experiments}
\input{040experiments}

\section{Related Work}
\label{sec:related}
\input{050related}

\section{Conclusion}
\label{sec:conclusion}
\input{060conclusion}

\section*{Acknowledgement}
\input{ACK-kijung}

\balance
\bibliographystyle{ACM-Reference-Format}
\bibliography{BIB/kijung}

\end{document}

%% file: dfn.tex
\newtheorem{problem}{Problem}

\newtheorem{property}{Property}

\newcommand{\hide}[1]{}

\newcommand{\tensor}[1]{\boldsymbol{\mathscr{#1}}}   
\newcommand{\mat}[1]{\mathbf{#1}}

\newcommand{\TR}{\tensor{R}}

\newcommand{\TT}{\tensor{T}}

\newcommand{\MA}{\mat{A}}

\newcommand{\bit}{\begin{compactitem}}
\newcommand{\eit}{\end{compactitem}}
\newcommand{\ben}{\begin{compactenum}}
\newcommand{\een}{\end{compactenum}}

\newcommand{\methodS}{\textsc{DenseStream}\xspace}
\newcommand{\methodA}{\textsc{DenseAlert}\xspace}
\newcommand{\dcube}{\textsc{D-Cube}\xspace}
\newcommand{\mzoom}{\textsc{M-Zoom}\xspace}
\newcommand{\cross}{\textsc{CrossSpot}\xspace}

\newcommand{\fraudar}{\textsc{Fraudar}\xspace}

\newcommand{\order}{D-ordering\xspace}
\newcommand{\aslice}{\TT_{(n,i_{n})}}
\newcommand{\aindex}{(i_{1},...,i_{N})}
\newcommand{\avalue}{t_{i_{1}...i_{N}}}
\newcommand{\increment}{(\aindex,\delta, +)}
\newcommand{\decrement}{(\aindex,\delta, -)}
\newcommand{\cmax}{c_{max}}

\newcommand{\aset}{Q}
\newcommand{\amem}{q}
\newcommand{\amemtwo}{r}
\newcommand{\asubset}{\aset_{\pi,\amem}}

\newcommand{\asubsetopttwo}{\aset_{\pi,\amem^{*}}}
\newcommand{\asubtensor}{\TT(\asubset)}

\newcommand{\asubsetmax}{S_{max}}
\newcommand{\asubsetopt}{S^{*}}
\newcommand{\asimplesubtensor}{\TT(S)}
\newcommand{\asubtensormax}{\TT(\asubsetmax)}
\newcommand{\asubtensoropt}{\TT(\asubsetopt)}
\newcommand{\pifunnoarg}{\pi(\cdot)}
\newcommand{\dfunnoarg}{d_{\pi}(\cdot)}
\newcommand{\cfunnoarg}{c_{\pi}(\cdot)}
\newcommand{\pifun}[1]{\pi(#1)}
\newcommand{\pifunnew}[1]{\pi'(#1)}
\newcommand{\piinvfun}[1]{\pi^{-1}(#1)}
\newcommand{\dfun}[1]{d(#1)}
\newcommand{\dpifun}[1]{d_{\pi}(#1)}
\newcommand{\cpifun}[1]{c_{\pi}(#1)}
\newcommand{\adegree}{\dpifun{\amem}}
\newcommand{\acore}{\cpifun{\amem}}
\newcommand{\asum}[1]{sum(#1)}
\newcommand{\annz}[1]{nnz(#1)}
\newcommand{\asimpledegree}{\dfun{\asimplesubtensor, \amem}}

\newcommand{\density}[1]{\rho(#1)}
\newcommand{\densityopt}{\rho_{opt}}
\newcommand{\densitynoarg}{\rho}

\DeclareMathOperator*{\argmax}{arg\,max}
\DeclareMathOperator*{\argmin}{arg\,min}

\newcommand{\cmark}{\ding{51}}%

\newcommand{\bluecolor}{\textcolor{blue}}

%% file: 000abstract.tex
Consider a stream of retweet events -
how can we spot fraudulent lock-step behavior in such multi-aspect data (i.e., tensors) evolving over time?
Can we detect it in real time, with an accuracy guarantee?
Past studies have shown that dense subtensors tend to indicate anomalous or even fraudulent behavior in many tensor data, including social media, Wikipedia, and TCP dumps.
Thus, several algorithms 
have been proposed for detecting dense subtensors rapidly and accurately.
However, existing algorithms assume that tensors are static,
while many real-world tensors, including those mentioned above, evolve over time. 

We propose \methodS, an incremental algorithm that maintains and updates a dense subtensor in a tensor stream
(i.e., a sequence of changes in a tensor), and \methodA, an incremental algorithm spotting the sudden appearances of dense subtensors.
Our algorithms are:
(1) {\bf Fast and `any time'}: updates by our algorithms are up to \textbf{a million times faster} 
than the fastest batch algorithms,
(2) {\bf Provably accurate}: our algorithms guarantee a lower bound on the density of the subtensor they maintain,
and (3) {\bf Effective}: our \methodA successfully spots anomalies in real-world tensors, especially those overlooked by existing algorithms.

%% file: 010intro.tex
\begin{figure*}[t]
	\centering
	\subfigure[Speed and accuracy of \methodS]{\label{fig:crown:tradeoff}
		\includegraphics[width=0.01\linewidth]{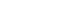}
		\includegraphics[width=0.305\linewidth]{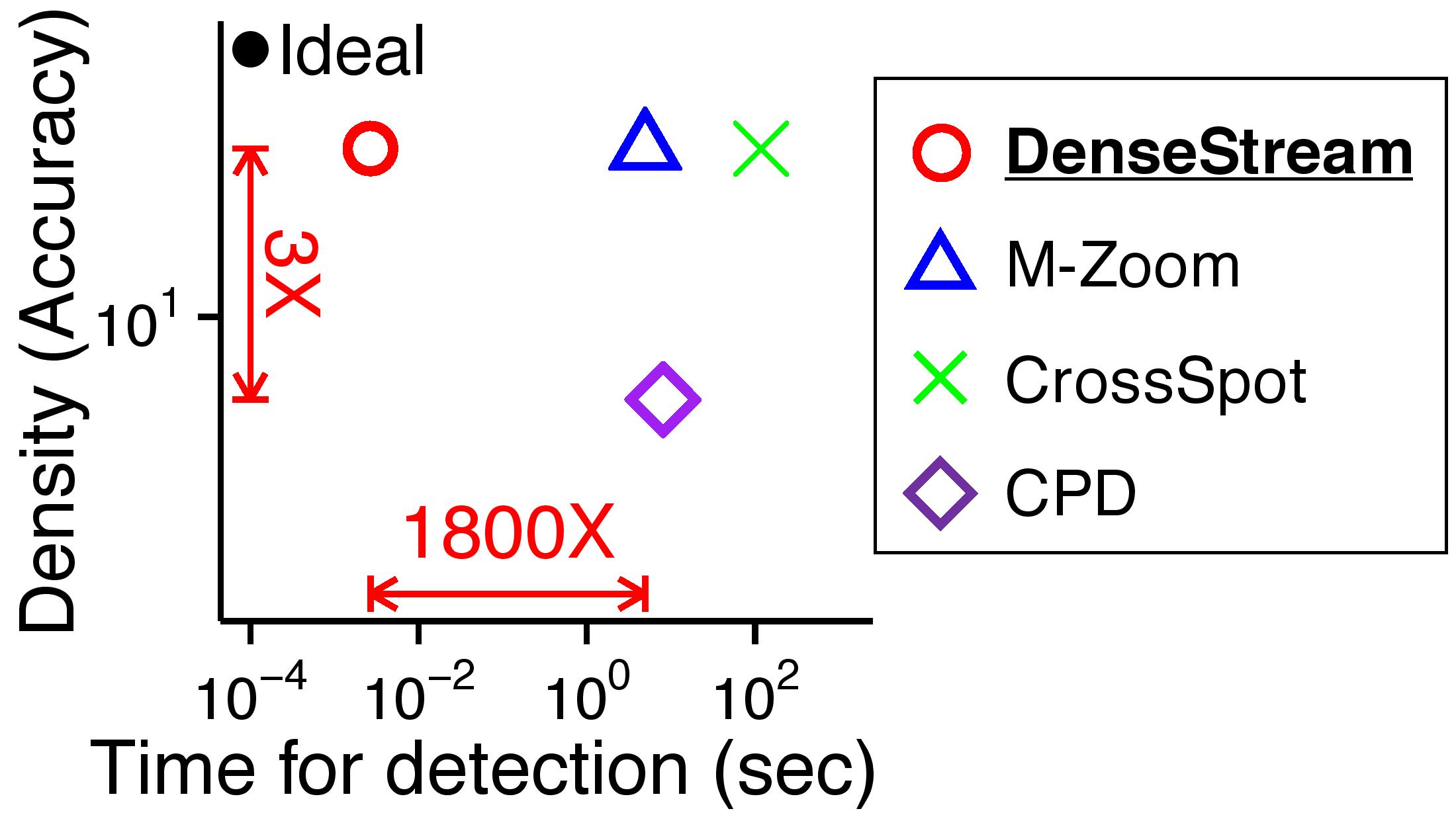}
		\includegraphics[width=0.01\linewidth]{FIG/space}
	}
	\subfigure[Scalability of \methodS]{\label{fig:crown:scalable}
		\includegraphics[width=0.33\linewidth]{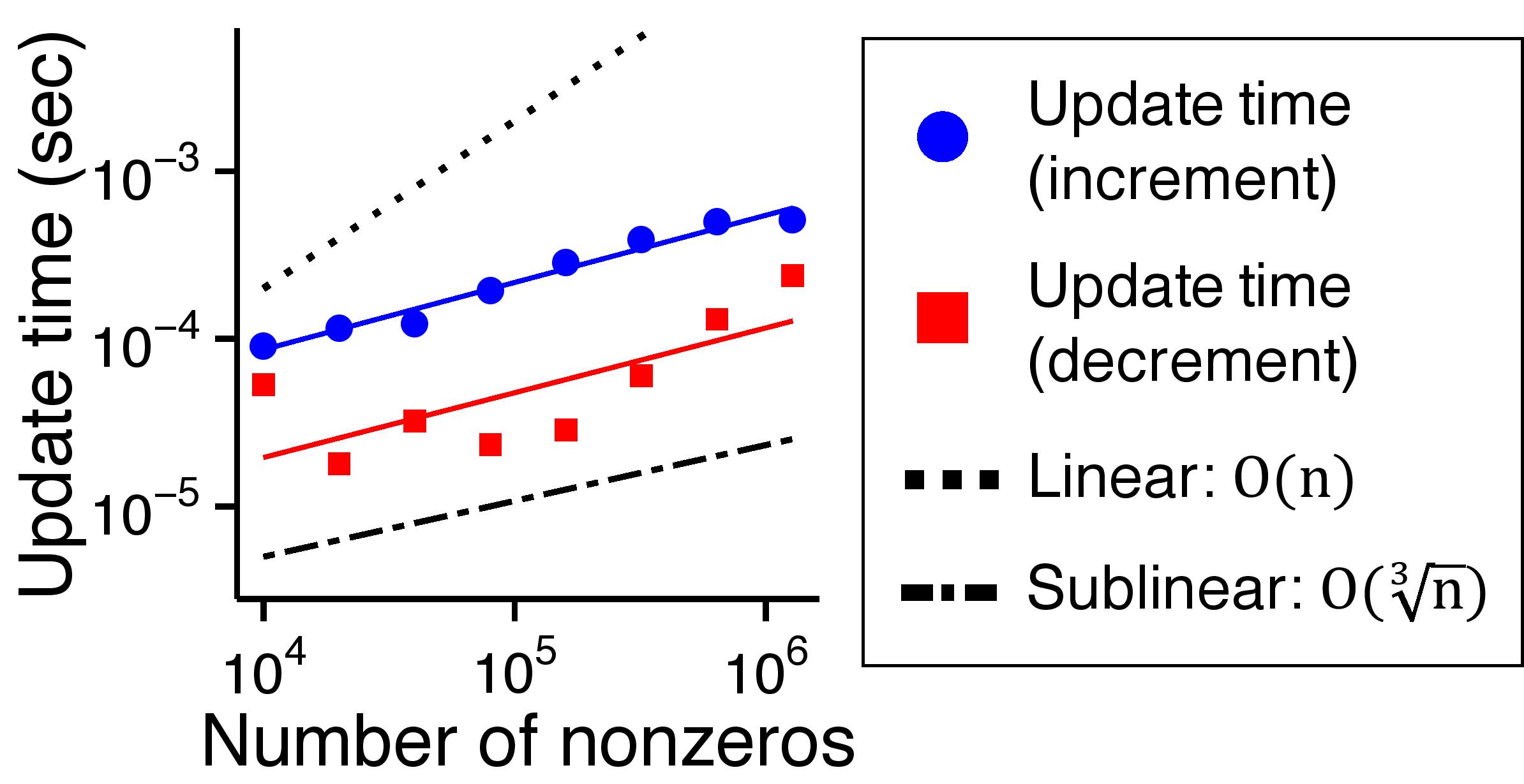}
	}
	\subfigure[Effectiveness of \methodA]{\label{fig:crown:tcp}
		\includegraphics[width=0.28\linewidth]{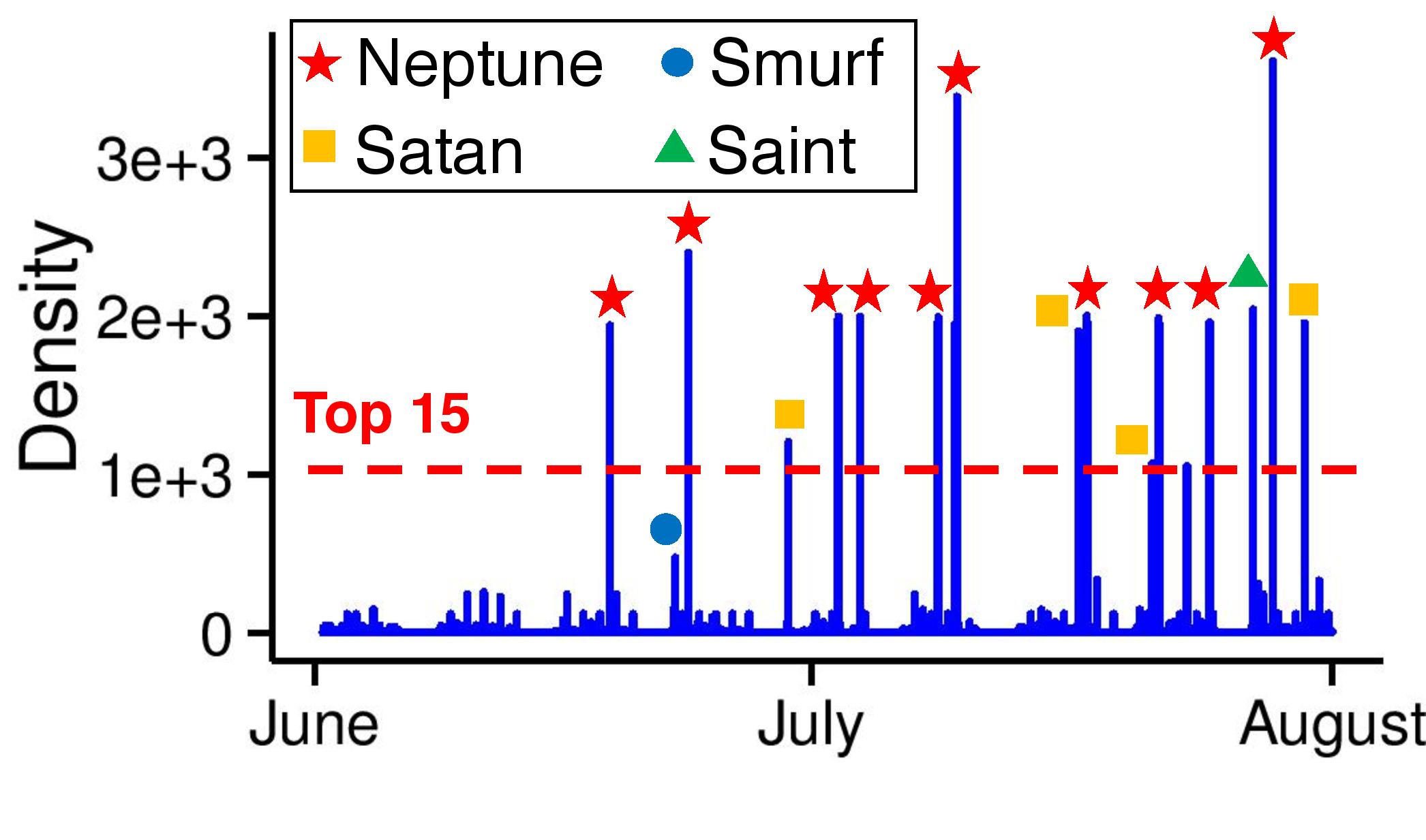}
	} 
	\caption{\label{fig:crown}
		\underline{\smash{Proposed algorithms are fast, accurate, scalable, and effective.}}
		(a) \methodS, our incremental algorithm, detects dense subtensors significantly faster than batch algorithms without losing accuracy, as seen in the result in Yelp Dataset.
		(b) The time taken for each update in \methodS grows sub-linearly with the size of data.
		(c) \methodA, which detects suddenly emerging dense subtensors, identifies network attacks from a TCP Dump with high accuracy (AUC=0.924). Especially, all the 15 densest subtensors revealed by \methodA indicate actual network attacks of various types.
	}
\end{figure*}

Given a stream of changes in a tensor that evolves over time, how can we detect the sudden appearances of dense subtensors? 

An important application of this problem is intrusion detection systems in networks, 
where attackers make a large number of connections to target machines to block their availability or to look for vulnerabilities~\cite{lippmann2000evaluating}. Consider a stream of connections where we represent each connection from a source IP address to a destination IP address as an entry in a 3-way tensor (source IP, destination IP, timestamp). Sudden appearances of dense subtensors in the tensor often indicate network attacks.
For example, in Figure~\ref{fig:crown:tcp}, all the top $15$ densest subtensors concentrated in a short period of time, which are detected by our \methodA algorithm, actually come from network attacks.

\begin{table}[h]
	\centering{
		\small
		\caption{\label{tab:compare} Comparison of \methodS, \methodA, and previous algorithms for detecting dense subtensors.}
		\begin{tabular}{c||ccccc|cc}
			\toprule
			\ &  \rotatebox[origin=l]{90}{{\mzoom \cite{shin2016mzoom}}} & 
			\rotatebox[origin=l]{90}{{\dcube \cite{shin2017dcube}}} &   
			\rotatebox[origin=l]{90}{{\cross \cite{jiang2015general}}}
			& \rotatebox[origin=l]{90}{{MAF \cite{maruhashi2011multiaspectforensics}}}
			& \rotatebox[origin=l]{90}{{\fraudar \cite{hooi2016fraudar}}}  & \rotatebox[origin=l]{90}{\methodS}
			& \rotatebox[origin=l]{90}{\methodA}  \\
			\midrule
			Multi-Aspect Data  & \cmark & \cmark &  \cmark & \cmark  &  & {\large \cmark} & {\large \cmark} \\
			Accuracy Guarantees & \cmark &  \cmark & & & \cmark & {\large \cmark} & {\large \cmark} \\
			Incremental Updates & &  & &  &  & {\large \cmark} & {\large \cmark} \\
			Slowly Formed Dense Subtensors & \cmark  &  \cmark &  \cmark  & \cmark  & \cmark  & {\large \cmark} &  \\
			Small Sudden Dense Subtensors &  & &  & & & & {\large \cmark}  \\
			\bottomrule
		\end{tabular}
	}
\end{table}

Another application is detecting fake rating attacks in review sites, such as Amazon and Yelp. Ratings can be modeled as entries in a 4-way tensor (user, item, timestamp, rating). Injection attacks maliciously manipulate the ratings of a set of items by adding a large number of similar ratings for the items, creating dense subtensors in the tensor. To guard against such fraud, an alert system detecting suddenly appearing dense subtensors in real time, as they arrive, is desirable.

Several algorithms for dense-subtensor detection have been proposed for detecting network attacks \cite{maruhashi2011multiaspectforensics, shin2016mzoom,shin2017dcube}, retweet boosting \cite{jiang2015general}, rating attacks \cite{shin2017dcube}, and bots \cite{shin2016mzoom} as well as for genetics applications \cite{saha2010dense}. As summarized in Table~\ref{tab:compare}, however, existing algorithms assume a static tensor rather than a stream of events (i.e., changes in a tensor) over time.
In addition, our experiments in Section~\ref{sec:experiments} show that they are limited in their ability to detect dense subtensors small but highly concentrated in a short period of time.

Our incremental algorithm \methodS detects dense subtensors in real time as events arrive, and is hence more useful in many practical settings, including those mentioned above.
\methodS is also used as a building block of \methodA, an incremental algorithm for detecting the sudden emergences of dense subtensors.
\methodA takes into account the tendency for lock-step behavior, such as network attacks and rating manipulation attacks, to appear within short, continuous intervals of time, which is an important signal for spotting lockstep behavior. 

As the entries of a tensor change, our algorithms work by maintaining a small subset of subtensors that always includes a dense subtensor with a theoretical guarantee on its density.
By focusing on this subset, our algorithms detect a dense subtensor in a time-evolving tensor up to a million times faster than the fastest batch algorithms, while providing the same theoretical guarantee on the density of the detected subtensor.


In summary, the main advantages of our algorithms are:
\bit
	\item {\bf Fast and `any time'}:  incremental updates by our algorithms are up to {\em a million times faster} than the fastest batch algorithms (Figure~\ref{fig:crown:tradeoff}).
	\item {\bf Provably accurate}: our algorithms maintain a subtensor with a theoretical guarantee on its density, and in practice, its density is similar to that of subtensors found by the best batch algorithms (Figure~\ref{fig:crown:tradeoff}).
	\item {\bf Effective}: \methodA successfully detects bot activities and network intrusions (Figure~\ref{fig:crown:tcp}) in real-world tensors. It also spots small-scale rating manipulation attacks, overlooked by existing algorithms.
\eit
{\bf Reproducibility}: The code and data
we used in the paper are available at \url{http://www.cs.cmu.edu/~kijungs/codes/alert}.

In Section~\ref{sec:prelim}, we introduce notations and problem
definitions. 
In Section~\ref{sec:method}, we describe our proposed algorithms: \methodS and \methodA.
In Section~\ref{sec:experiments}, we present experimental
results. After reviewing related work in Section~\ref{sec:related}, 
we conclude in Section~\ref{sec:conclusion}.

%% file: 020prelim.tex
In this section, we introduce notations and concepts used in the paper.
Then, we give formal problem definitions.

\subsection{Notations and Concepts.}
\label{sec:prelim:notation}

Symbols frequently used in the paper are listed in Table~\ref{tab:symbols}, and a toy example is in Example~\ref{example}.
We use $[y]=\{1,2...,y\}$ for brevity.

\begin{table}[!t]
	\centering
	\small
	\caption{Table of symbols.}
	\begin{tabular}{c|l}
		\toprule
		\textbf{Symbol} & \textbf{\qquad\qquad\qquad Definition} \\
		\midrule
		$\TT$ & an input tensor\\
		$N$ & order of $\TT$ \\
		$\avalue$ & entry of $\TT$ with index $\aindex$ \\
		$\aset$ & set of the slice indices of $\TT$ \\
		$\amem$ & a member of $\aset$ \\
		$\asimplesubtensor$ & subtensor composed of the slices in $S\subset \aset$\\
		\midrule
		$\pi:[|\aset|] \rightarrow \aset$ & an ordering of slice indices in $\aset$ \\
		$\asubset$ & slice indices located after or equal to $\amem$ in $\pi$ \\
		\midrule
		$\asum{\asimplesubtensor}$ & sum of the entries included in $\asimplesubtensor$ \\
		$\asimpledegree$ & slice sum of $\amem$ in $\asimplesubtensor$ \\
		$\adegree$ & slice sum of $\amem$ in $\asubtensor$ \\
		$\acore$ & cumulative max. slice sum of $\amem$ in $\asubtensor$ \\
		\midrule
		$\increment$ & increment of $\avalue$ by $\delta$\\
		$\decrement$ & decrement of $\avalue$ by $\delta$ \\
		\midrule
		$\density{\asimplesubtensor}$ & density of a subtensor $\asimplesubtensor$ \\
		$\densityopt$ & density of the densest subtensor in $\TT$ \\
		\midrule
		$\Delta T$ & time window in \methodA \\
		$[y]$ & $\{1,2...,y\}$ \\
		\bottomrule
	\end{tabular}
	\label{tab:symbols}
\end{table}


{\bf \underline{Notations for Tensors}:}
Tensors are multi-dimensional arrays that generalize vectors (1-way tensors) and matrices (2-way tensors) to higher orders.
Consider an $N$-way tensor $\TT$ of size $I_{1} \times ... \times I_{N}$ with non-negative entries.
Each $\aindex$-th entry of $\TT$ is denoted by $\avalue$.
Equivalently, each {\it $n$-mode index} of $\avalue$ is $i_{n}$.
We use $\aslice$ to denote the {\it $n$-mode slice} (i.e. ($N-1$)-way tensor) obtained by fixing $n$-mode index to $i_{n}$.
Then, $\aset=\{(n,i_{n}):n\in[N], i_{n}\in[I_{n}]\}$ indicates all the slice indices.
We denote a member of $\aset$ by $\amem$.

For example, if $N=2$, $\TT$ is a matrix of size $I_{1}\times I_{2}$. Then, $\TT_{(1,i_{1})}$ is the $i_{1}$-th row of $\TT$, and $\TT_{(2,i_{2})}$ is the $i_{2}$-th column of $\TT$.
In this setting, $\aset$ is the set of all row and column indices.

{\bf \underline{Notations for Subtensors}:}
Let $S$ be a subset of $\aset$.
$\TT(S)$ denotes the subtensor composed of the slices with indices in $S$, i.e., $\TT(S)$ is the subtensor left after removing all the slices with indices not in $S$.

For example, if $\TT$ is a matrix (i.e., $N=2$) and $S=\{(1,1),(1,2),$ $(2,2),(2,3)\}$, $\TT(S)$ is the submatrix of $\TT$ composed of the first and second rows and the second and third columns.

{\bf \underline{\smash{Notations for Orderings}}:}
Consider an ordering of the slice indices in $\aset$.
A function $\pi:[|\aset|]\rightarrow \aset$ denotes such an ordering where, for each $j\in[|\aset|]$, $\pifun{j}$ is the slice index in the $j$th position.
That is, each slice index $\amem\in \aset$ is in the $\piinvfun{\amem}$-th position in $\pi$.
Let
$\asubset=\{\amemtwo \in \aset : \piinvfun{\amemtwo}\geq \piinvfun{\amem}\}$ be the slice indices located after or equal to $\amem$ in $\pi$.
Then, $\asubtensor$ is the subtensor of $\TT$ composed of the slices with their indices in $\asubset$.

{\bf \underline{\smash{Notations for Slice Sum}}:}
We denote the sum of the entries of $\TT$ included in subtensor $\asimplesubtensor$ by $\asum{\asimplesubtensor}$.
Similarly, we define {\it the slice sum} of $\amem\in \aset$ in subtensor $\asimplesubtensor$, denoted by $\asimpledegree$, as the sum of the entries of $\TT$ that are included in both $\asimplesubtensor$ and the slice with index $\amem\in \aset$.
For an ordering $\pi$ and a slice index $\amem \in \aset$, we use  $\adegree=\dfun{\asubtensor,\amem}$ for brevity, and define
the {\it cumulative maximum slice sum} of $\amem$ as $\acore = \max\{\dpifun{\amemtwo}: \amemtwo\in \aset, \piinvfun{\amemtwo} \leq \piinvfun{\amem}\}$, i.e., maximum $\dfunnoarg$ among the slice indices located before or equal to $\amem$ in $\pi$.

{\bf \underline{\smash{Notations for Tensor Streams}}:}
A tensor stream is a sequence of changes in $\TT$.
Let $\increment$ be an increment of entry $\avalue$ by $\delta>0$ and $\decrement$ be a decrement of entry $\avalue$ by $\delta>0$.

\begin{figure}[t]
	\centering
	\vspace{-2mm}
	\includegraphics[width=0.7\linewidth]{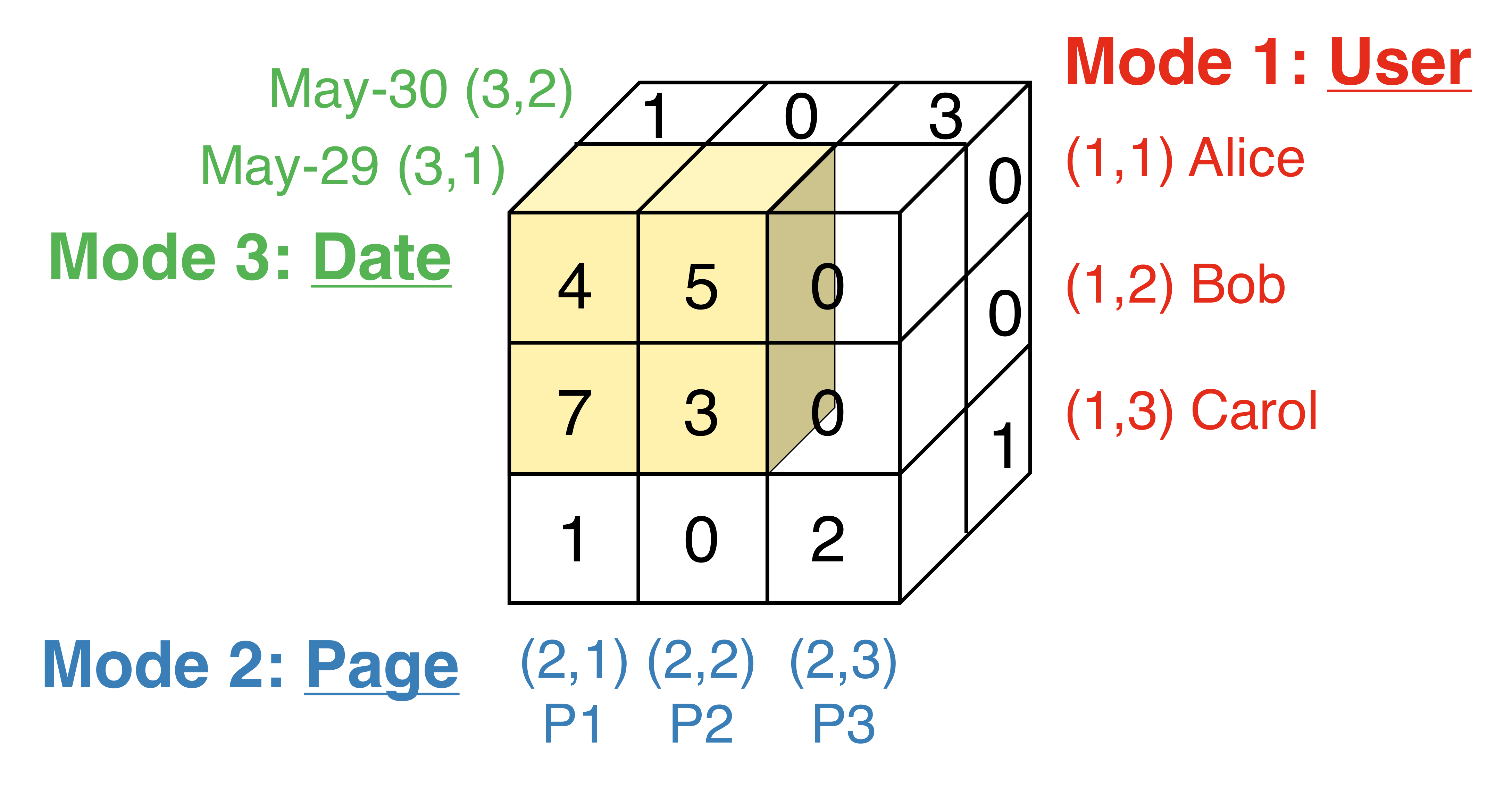}
	\vspace{-1mm}
	\caption{\label{fig:example}
		Pictorial depiction of Example~\ref{example}. The colored region indicates subtensor $\asimplesubtensor$.
	}
\end{figure}

\vspace{-1mm}
\begin{example}[Wikipedia Revision History] \label{example}
	Consider the $3$-way tensor in Figure~\ref{fig:example}.
	In the tensor, each entry $t_{ijk}$ indicates that user $i$ revised page $j$ on date $k$, $t_{ijk}$ times.
	The set of the slice indices is $\aset=\{(1,1),(1,2),(1,3),(2,1),(2,2),(2,3),$ $(3,1),(3,2)\}$.
	Consider its subset $S=\{(1,1),(1,2),(2,1),$ $(2,2),(3,1)\}$.
	Then, $\asimplesubtensor$ is the subtensor composed of the slices with their indices in $S$, as seen in Figure~\ref{fig:example}.
	In this setting, $\asum{\asimplesubtensor}=4+5+7+3=19$, and $\dfun{\asimplesubtensor,(2,2)}=5+3=8$.
	Let $\pi$ be an ordering of $\aset$ where $\pifun{1}=(1,3)$, 
	$\pifun{2}=(2,3)$, $\pifun{3}=(3,2)$, $\pifun{4}=(2,2)$, $\pifun{5}=(1,1)$, $\pifun{6}=(1,2)$, $\pifun{7}=(2,1)$, and $\pifun{8}=(3,1)$.
	Then, $\aset_{\pi,(2,2)}=S$, and $\dpifun{(2,2)}=\dfun{\TT(\aset_{\pi,(2,2)}),(2,2)}=\dfun{\asimplesubtensor,$ $(2,2)}=8$.
\end{example}

\vspace{-1mm}
\subsection{Density Measure.}
\label{sec:prelim:density}
Definition~\ref{defn:density} gives the density measure used in this work.
That is, the density of a subtensor is defined as the sum of its entries divided by the number of the slices composing it.
We let $\densityopt$ be the density of the densest subtensor in $\TT$.

\begin{definition}\label{defn:density} \textsc{(Density of a subtensor \cite{shin2016mzoom}).} {\it Consider a subtensor $\asimplesubtensor$ of a tensor $\TT$. The density of $\asimplesubtensor$, which is denoted by $\density{\asimplesubtensor}$, is defined as $$\density{\asimplesubtensor}=\frac{\asum{\asimplesubtensor}}{|S|}.$$
		}
\end{definition}

This measure is chosen because:
(a) it was successfully used for anomaly and fraud detection \cite{shin2016mzoom,shin2017dcube},
(b) this measure satisfies axioms that a reasonable ``anomalousness" measure should meet (see Section~A of the supplementary document \cite{supple}),
and (c) our algorithm based on this density measure outperforms existing algorithms based on different density measures in Section~\ref{sec:exp:effective:rating}.

\subsection{Problem Definitions.}
\label{sec:prelim:problem}

We give the formal definitions of the problems studied in this work. 
The first problem (Problem~\ref{problem:stream}) is to maintain the densest subtensor in a tensor that keeps changing.

\begin{problem}[Detecting the Densest Subtensor in a Tensor Stream]\label{problem:stream}
\textbf{(1) Given:} a sequence of changes in a tensor $\TT$ with slice indices $\aset$ (i.e., a tensor stream)
\textbf{(2) maintain:} a subtensor $\asimplesubtensor$ where $S\subset \aset$, \textbf{(3) to maximize:} its density $\density{\asimplesubtensor}$.
\end{problem}

Identifying the exact densest subtensor is 
computationally expensive even for a static tensor. 
For example, it takes $O(|\aset|^{6})$ even when $\TT$ is a binary matrix (i.e., $N=2$) \cite{goldberg1984finding}.
Thus, we focus on designing an approximation algorithm that maintains a dense subtensor with a provable approximation bound, significantly faster than repeatedly finding a dense subtensor from scratch.

The second problem (Problem~\ref{problem:time}) is to detect suddenly emerging dense subtensors in a tensor stream.
For a tensor $\TT$ whose values increase over time, let $\TT_{\Delta T}$ be the tensor where the value of each entry is the increment in the corresponding entry of $\TT$ in the last $\Delta T$ time units.
Our aim is to spot dense subtensors appearing in $\TT_{\Delta T}$, which also keeps changing.

\begin{problem}[Detecting Sudden Dense Subtensors in a Tensor Stream]\label{problem:time}
	\textbf{(1) Given:} a sequence of increments in a tensor $\TT$ with slice indices $\aset$ (i.e., a tensor stream) and a time window $\Delta T$,
	\textbf{(2) maintain:} a subtensor $\TT_{\Delta T}(S)$ where $S\subset \aset$, \textbf{(3) to maximize:} its density $\density{\TT_{\Delta T}(S)}$.
\end{problem}


%% file: 030method.tex

In this section, we propose \methodS, which is an incremental algorithm for dense-subtensor detection in a tensor stream, and \methodA, which detects suddenly emerging dense subtensors.
We first explain dense-subtensor detection in a static tensor  in Section~\ref{sec:method:static}, then generalize this to \methodS for a dynamic tensor in Section~\ref{sec:method:dynamic}.
Finally, we propose \methodA based on \methodS in Section~\ref{sec:method:alert}.

\begin{algorithm}[t]
	\small
	\caption{Dense-subtensor detection in a static tensor}\label{alg:static}
	\begin{algorithmic}[1] 
		\Require a tensor $\TT$ with slice indices $Q$
		\Ensure a dense subtensor $\asubtensormax$
		
		\State compute $\pifunnoarg$, $\dfunnoarg$, $\cfunnoarg$ by \Call{\order}{}()
		\State $\asubsetmax\leftarrow$ \Call{Find-Slices}{}()
		\State \textbf{return} $\asubtensormax$ 
		
		\Procedure{\order}{}():
		\Statex \hfill \bluecolor{$\vartriangleright$ find a \order $\pifunnoarg$ and compute $\dfunnoarg$ and $\cfunnoarg$}
		\State $S \gets \aset$; \ $\cmax \gets 0$   \hfill \bluecolor{$\vartriangleright$ $\cmax$: max. $\dfunnoarg$ so far}
		\For{$j\leftarrow 1...|\aset|$} 
		\State $\amem \gets \argmin_{\amemtwo \in S} \dfun{\asimplesubtensor, \amemtwo}$ \label{alg:static:line:minslice} \hfill \bluecolor{$\vartriangleright$ $q$ has min. slice sum} 
		\State $\pifun{j} \gets \amem$ \hfill \bluecolor{$\vartriangleright$ $S=\asubset$}
		\State $\adegree \gets \asimpledegree$ \hfill \bluecolor{$\vartriangleright$ $\adegree=\dfun{\asubtensor, \amem}$}
		\State $\acore \gets \max(\cmax, \adegree)$; \ $\cmax \gets \acore $
		\State $S \gets S/\{\amem\}$ 
		\EndFor
		\EndProcedure
		
		\Procedure{Find-Slices}{}():
		\Statex \hfill \bluecolor{$\vartriangleright$ find slices forming a dense subtensor from $\pifunnoarg$ $\dfunnoarg$, and $\cfunnoarg$}
		\State $S \gets \emptyset$; \ $m \gets 0$; \hfill  \bluecolor{$\vartriangleright$ $m$: $\asum{\asimplesubtensor}$}
		\State $\densitynoarg_{max} \gets -\infty$; \ $q_{max} \gets 0$ \hfill 
		\bluecolor{$\vartriangleright$ $\densitynoarg_{max}$: max. density so far}
		\For{$j\leftarrow |\aset|..1$} 
		\State $\amem \gets \pi(j);\ S \gets S\cup \{\amem\}$ \hfill \bluecolor{$\vartriangleright$ $S=\asubset$}
		\State $m \gets m + \adegree$ \hfill \bluecolor{$\vartriangleright$ $m=\asum{\asubtensor}$}
		\If{$m/|S| > \densitynoarg_{max}$} 
		\vspace{-3.4mm}
		\Statex \hfill \bluecolor{$\vartriangleright$ $m/|S|=\density{\asubtensor}$}
		\State $\densitynoarg_{max} \gets m/|S|$; \ $q_{max} \gets q$
		\EndIf
		\EndFor
		\State \textbf{return} $\aset_{\pi, q_{max}}$ \hfill
		 \bluecolor{$\vartriangleright$ $q_{max}=\argmax_{\amem\in \aset }\density{\asubtensor}$}
		\EndProcedure		
	\end{algorithmic}
\end{algorithm}

\subsection{Dense Subtensor Detection in Static Data.}
\label{sec:method:static}

We propose Algorithm~\ref{alg:static} for detecting a dense subtensor in a static tensor.
Although it eventually finds the same subtensor as \mzoom \cite{shin2016mzoom}, Algorithm~\ref{alg:static} also computes extra information, including a \textit{\order} (Definition~\ref{defn:core_order}), required for updating the subtensor in the following sections.
Algorithm~\ref{alg:static} has two parts: (a) \textbf{\order}: find a \order $\pi$ and compute $\dfunnoarg$ and $\cfunnoarg$; and (b) \textbf{Find-Slices}: find slices forming a dense subtensor from the result of (a).

\begin{definition} \label{defn:core_order} \textsc{(\order).} {\it 
	An ordering $\pi$ is a {\bf \order} of $\aset$ in $\TT$ if $\forall \amem \in \aset$, $
	\dfun{\asubtensor,q}=\min_{\amemtwo \in \asubset}\dfun{\asubtensor, \amemtwo }$.
	}
\end{definition}
That is, a \order is an ordering of slice indices obtained by choosing a slice index with minimum slice sum repeatedly, as in \textsc{\order}\textsc{()} of Algorithm~\ref{alg:static}.

Using a \order drastically reduces the search space while providing a guarantee on the accuracy.
With a \order $\pi$, Algorithm~\ref{alg:static} reduces the search space of $2^{|\aset|}$ possible subtensors to $\{\asubtensor:q\in \aset\}$.
In this space of size $|\aset|$, however, there always exists a subtensor whose density is at least 1/(order of the input tensor) of maximum density, as formalized in Lemmas~\ref{lemma:maxdensity} and \ref{lemma:static:order}.

\begin{lemma} \label{lemma:maxdensity}
	Let $\asubtensoropt$ be a subtensor with the maximum density, i.e., $\density{\asubtensoropt} = \densityopt$. Then for any $\amem \in \asubsetopt$,
	\begin{equation}
	d(\asubtensoropt,\amem)\geq\density{\asubtensoropt}.\label{eq:maxblockDegreeDensity}
	\end{equation} 
	\noindent {\it Proof.} The maximality of the density of $\asubtensoropt$ implies \\
	$\density{\TT(\asubsetopt \backslash\{\amem\})}\leq\density{\asubtensoropt}$,
	and plugging in Definition 2.2 to $\rho$ gives 
	\begin{multline*}
		\frac{\asum{\asubtensoropt}-d(\asubtensoropt,\amem)}{|\asubsetopt|-1}=\frac{\asum{\TT(\asubsetopt\backslash\{\amem\})}}{|\asubsetopt|-1} \\ =\density{\TT(\asubsetopt\backslash\{\amem\})}
		\leq\density{\asubtensoropt}=\frac{\asum{\asubtensoropt}}{|\asubsetopt|},
	\end{multline*}
	which reduces to Eq.~\eqref{eq:maxblockDegreeDensity}. \qed
\end{lemma}

\begin{lemma} \label{lemma:static:order}
	Given a \order $\pi$ in an $N$-way tensor $\TT$, there exists $\amem\in \aset$ such that $\density{\asubtensor} \geq \densityopt/N$. \vspace{0.5mm} \\
	\noindent {\it Proof.} Let $\asubtensoropt$ be satisfying $\density{\asubtensoropt}=\densityopt$, and
	let $\amem^{*}\in \asubsetopt$ be satisfying
	that $\forall \amem\in \asubsetopt$, $\pi^{-1}(\amem^{*})\leq\piinvfun{\amem}$. Our
	goal is to show $\density{\TT(\asubsetopttwo)}\geq\frac{1}{N}\density{\asubtensoropt}$,
	which we show as $N\density{\TT(\asubsetopttwo)}\geq d(\TT(\asubsetopttwo),\amem^{*})\geq d(\asubtensoropt,\amem^{*})\geq\density{\asubtensoropt}$.
	
	To show $N\density{\TT(\asubsetopttwo)}\geq d(\TT(\asubsetopttwo),\amem^{*})$,
	note  $N\density{\TT(\asubsetopttwo)}=\frac{\asum{\TT(\asubsetopttwo)}N}{\left|\asubsetopttwo\right|}$,
	and since $\TT$ is an $N$-way tensor, each entry is included in $N$ slices. Hence 
	\begin{equation}
	\sum\nolimits_{\amem\in \asubsetopttwo}d(\TT(\asubsetopttwo),\amem)=\asum{\TT(\asubsetopttwo)}N.\label{eq:maxdensity_degreesum}
	\end{equation}
	Since $\pi$ is a $D$-ordering, $\forall\amem\in \asubsetopttwo$,
	$d(\TT(\asubsetopttwo),\amem)$ $\geq d(\TT(\asubsetopttwo),\amem^{*})$
	holds. Combining this and Eq. \eqref{eq:maxdensity_degreesum} gives 
	\begin{align*}
		 N\density{\TT(\asubsetopttwo)}& =\frac{\asum{\TT(\asubsetopttwo)}N}{\left|\asubsetopttwo\right|}\\
		& =\frac{\sum_{q\in \asubsetopttwo}d(\TT(\asubsetopttwo),\amem)}{\left|\asubsetopttwo\right|}\geq d(\TT(\asubsetopttwo),\amem^{*}).
	\end{align*}
	Second, $d(\TT(\asubsetopttwo),\amem^{*})\geq d(\asubtensoropt,\amem^{*})$
	is from that $\asubsetopt \subset\asubsetopttwo$.
	Third, $d(\asubtensoropt,\amem^{*})\geq\density{\asubtensoropt}$
	is from Lemma \ref{lemma:maxdensity}. From these, $\density{\TT(\asubsetopttwo)}\geq\frac{1}{N}\density{\asubtensoropt}$
	holds.	\qed
\end{lemma}

Such a subtensor $\asubtensormax$ is detected by Algorithm~\ref{alg:static}.
That is, $\asubtensormax$ has density at least 1/(order of the input tensor) of maximum density, as proved in Theorem~\ref{thm:static:accuracy}.

\begin{theorem}[Accuracy Guarantee of Algorithm~\ref{alg:static}] \label{thm:static:accuracy} 
	The subtensor returned by Algorithm~\ref{alg:static} has density at least $\densityopt/N$. \vspace{0.5mm} \\
	\noindent {\it Proof.}
	By Lemma~\ref{lemma:static:order}, there exists a subtensor with density at least $\densityopt/N$ among $\{\asubtensor:\amem\in \aset \}$.
	The subtensor with the highest density in the set is returned by Algorithm~\ref{alg:static}. \qed 
\end{theorem}

The time complexity of Algorithm~\ref{alg:static} is linear with $\annz{\TT}$, the number of the non-zero entries in $\TT$, as formalized in Theorem~\ref{thm:static:time}. Especially, finding $\asubsetmax$ takes only $O(|Q|)$ given $\pifunnoarg$, $\dfunnoarg$, and $\cfunnoarg$, as shown in Lemma~\ref{lemma:time:find}.

\begin{lemma}\label{lemma:time:find}
	Let $\asubsetmax$ be the set of slice indices returned by \textsc{Find-Slices()} in Algorithm~\ref{alg:static}, and let $\TT(\amem)$ be the set of the non-zero entries in the slice with index $q$ in $\TT$.
	The time complexity of \textsc{Find-Slices()} in Algorithm~\ref{alg:static} is $O(|\aset|)$ and that of constructing $\asubtensormax$ from $\asubsetmax$ is $O(N|\bigcup_{\amem\in \asubsetmax}\TT(\amem)|)$.
	\vspace{0.5mm} \\
	\noindent {\it Proof.}
	Assume that, for each slice, the list of the non-zero entries in the slice is stored.
	In \textsc{Find-Slices()}, we iterate over the slices in $\aset$, and each iteration takes $O(1)$. Thus, we get $O(|\aset|)$.
	After finding $\asubsetmax$, 
	in order to construct $\TT(\asubsetmax)$, we have to process each non-zero entry included in  any slice in $\asubsetmax$. The number of such entries is $|\bigcup_{\amem\in \asubsetmax}\TT(\amem)|$.
	Since processing each entry takes $O(N)$, constructing $\TT(\asubsetmax)$ takes $O(N|\bigcup_{\amem\in \asubsetmax}\TT(\amem)|)$. \qed
\end{lemma}

\begin{theorem}[Time Complexity of Algorithm~\ref{alg:static}] \label{thm:static:time}
	The time complexity of Algorithm~\ref{alg:static} is $O(|\aset|\log|\aset|+\annz{\TT}N)$.
	\vspace{0.5mm} \\
	\noindent {\it Proof.} Assume that, for each slice, the list of the non-zero entries in the slice is stored.
	We first show that the time complexity of \textsc{D-ordering()} in Algorithm~\ref{alg:static} is $O(|\aset|\log|\aset|+\annz{\TT}N|)$.
	Assume we use a Fibonacci heap to find slices with minimum slice sum (line~\ref{alg:static:line:minslice}).
	Computing the slice sum of every slice takes $O(\annz{\TT}N)$, and
	constructing a Fibonacci heap where each value is a slice index in $\aset$ and the corresponding key is the slice sum of the slice takes $O(|\aset|)$.
	Popping the index of a slice with minimum slice sum, which takes $O(\log|\aset|)$, happens $|\aset|$ times, and thus we get $O(|\aset|\log|\aset|)$.
	Whenever a slice index is popped we have to update the slice sums of its dependent slices (two slices are dependent if they have common non-zero entries). 
	Updating the slice sum of each dependent slice, which takes $O(1)$ in a Fibonacci heap, happens at most $O(\annz{\TT}N)$ times, and thus we get $O(\annz{\TT}N)$.
	Their sum results in $O(|\aset|\log|\aset|+\annz{\TT}N)$.
	
	By Lemma~\ref{lemma:time:find}, the time complexity of \textsc{Find-Slices()} is $O(|\aset|)$, and that of constructing $\asubtensormax$ from $\asubsetmax$ is $O(N|\bigcup_{\amem\in \asubsetmax}\TT(\amem)|).$
	
	Since the time complexity of \textsc{D-ordering()} dominates that of the remaining parts, we get $O(|\aset|\log|\aset|+\annz{\TT}N)$ as the time complexity of Algorithm~\ref{alg:static}. \qed
\end{theorem}

\subsection{\textsc{\Large \methodS:} Dense-Subtensor Detection in a Tensor Stream.}\label{sec:method:dynamic}

How can we update the subtensor found in Algorithm~\ref{alg:static} under changes in the input tensor, rapidly, only when necessary, with the same approximation bound?
For this purpose, we propose \methodS, which updates the subtensor while satisfying Property~\ref{property:dynamic}. We explain the responses of \methodS to increments of entry values (Section~\ref{sec:method:dynamic:inc}), decrements of entry values (Section~\ref{sec:method:dynamic:dec}), and changes of the size of the input tensor (Section~\ref{sec:method:dynamic:size}).

\begin{property} [Invariants in \methodS]\label{property:dynamic}
	For an $N$-way tensor $\TT$ that keeps changing, the ordering $\pi$ of the slice indices and the dense subtensor $\density{\asubtensormax}$ maintained by \methodS satisfy the following two conditions:
	\bit
	\item $\pi$ is a \order of $\aset$ in $\TT$
	\item $\density{\asubtensormax}$ $\geq\densityopt/N$.
	\eit
\end{property}

\subsubsection{Increment of Entry Values.}
\label{sec:method:dynamic:inc}
Assume that the maintained dense subtensor $\asubtensormax$ and ordering $\pi$ (with $\dfunnoarg$ and $\cfunnoarg$) satisfy Property~\ref{property:dynamic} in the current tensor $\TT$ (such $\pi$, $\dfunnoarg$, $\cfunnoarg$, and $\asubtensormax$ can be initialized by Algorithm~\ref{alg:static} if we start from scratch).
Algorithm~\ref{alg:method:increment} describes the response of
\methodS to $\increment$, an increment of entry $\avalue$ by $\delta >0$, for satisfying Property~\ref{property:dynamic}.
Algorithm~\ref{alg:method:increment} has three steps: (a) \textbf{Find-Region}: find a region of the \order $\pi$ that needs to be reordered, (b) \textbf{Reorder}: reorder the region obtained in (a), and (c) \textbf{Update-Subtensor}: use $\pi$ to rapidly update $\asubtensormax$ only when necessary. Each step is explained below.

{\bf (a) Find-Region (Line~\ref{alg:method:increment:line:find}):} The goal of this step is to find the region $[j_{L}, j_{H}] \subset [1, |Q|]$ of the domain of the \order $\pi$ that needs to be reordered in order for $\pi$ to remain as a \order after the change $\increment$.
Let  $C=\{(n,i_{n}): n\in[N]\}$ be the indices of the slices composing the changed entry $\avalue$ and let $\amem_{f}=\argmin_{\amem\in C}\piinvfun{\amem}$ be the one located first in $\pi$ among $C$.
Then, let $M=\{\amem \in \aset : \piinvfun{\amem} > \piinvfun{\amem_{f}}, \ \adegree \geq \dpifun{\amem_{f}} + \delta  \}$ be the slice indices that are located after $\amem_{f}$ in $\pi$ among $Q$ and having $d_{\pi}(\cdot)$ at least $d_{\pi}(\amem_{f})+\delta$.
Then, $j_{L}$ and $j_{H}$ are set as follows:
\begin{align}
j_{L}= &  \piinvfun{\amem_{f}}, \label{eq:increment:min}\\ 
j_{H}= & \begin{cases}
	\min_{\amem \in M} \piinvfun{\amem}-1 & \text{if}\ M \neq \emptyset, \\
	|\aset|  \text{ (i.e., the last index)} & \text{otherwise.} \label{eq:increment:max}
\end{cases}
\end{align}

Later in this section, we prove that slice indices whose locations do not belong to $[j_{L},j_{H}]$ need not be reordered by showing that there always exists a \order $\pi'$ in the updated $\TT$ where $\pifunnew{j}=\pifun{j}$ for every $j\notin[j_{L},j_{H}]$. 

\begin{algorithm}[t]
	\small
	\caption{\methodS in the case of increment}\label{alg:method:increment}
	\begin{algorithmic}[1] 
		\Require (1) current tensor: $\TT$ with slice indices $Q$
		\Statex \ \ \quad (2) current dense subtensor: $\asubtensormax$ with Property~\ref{property:dynamic}
		\Statex \ \ \quad (3) current \order: $\pifunnoarg$ (also $\dfunnoarg$ and $\cfunnoarg$)
		\Statex \ \ \quad (4) a change in $\TT$: $\increment$
		\Ensure updated dense subtensor $\asubtensormax$
		\State $\avalue \gets \avalue + \delta$ 
		\State compute $j_{L}$ and $j_{H}$ by Eq. \eqref{eq:increment:min} and Eq.~\eqref{eq:increment:max} \label{alg:method:increment:line:find} \hfill \bluecolor{$\vartriangleright$  \textsc{[Find-Region]}}
		\State compute $R$ by Eq.~\eqref{eq:increment:R} \label{alg:method:increment:line:reorder:start} \hfill \bluecolor{$\vartriangleright$ \textsc{[Reorder]}}
		\State $S \gets \{\amem \in \aset  : \piinvfun{\amem} \geq j_{L}\}$;  $RS \leftarrow R\cap S$
		\State $c_{max} \gets 0$ \hfill \bluecolor{$\vartriangleright$ $\cmax$: max. $\dfunnoarg$ so far}
		\State \textbf{if} $j_{L}>1$ \textbf{then} $c_{max} \gets c_{\pi}(\pi(j_{L}-1))$ 
		\For{$j\leftarrow j_{L}...j_{H}$} 
		\State $\amem \gets \argmin_{\amemtwo \in RS} \dfun{\asimplesubtensor, \amemtwo}$ \label{alg:increment:line:minslice} \hfill \bluecolor{$\vartriangleright$ $q$ has min. slice sum}
		\State $\pifun{j} \gets \amem$ \hfill \bluecolor{$\vartriangleright$ by Lemma~\ref{lemma:increment:order}, $S=\asubset$, $RS= R \cap \asubset$}
		\State $\adegree \gets \asimpledegree$ \hfill \bluecolor{$\vartriangleright$ $\adegree=\dfun{\asubtensor, \amem}$}
		\State $\acore \gets \max(\cmax, \adegree)$; \ $\cmax\gets \acore$
		\State $S \gets S/\{\amem\}$; $RS \gets RS/\{\amem\}$ \label{alg:method:increment:line:reorder:end}
		\EndFor
	
		\If{$c_{max} \geq \density{\asubtensormax}$} \label{alg:method:increment:line:cond} \label{alg:method:increment:line:update:start} \hfill \bluecolor{$\vartriangleright$  \textsc{[Update-Subtensor]}}
		\State $S'\gets$\Call{Find-Slices}{}() in Algorithm~\ref{alg:static} \hfill \bluecolor{$\vartriangleright$ time complexity: $O(|Q|)$} 
		\If{$\asubsetmax \neq S'$}{  $\asubtensormax\gets\TT(S')$} \EndIf
		\EndIf \label{alg:method:increment:line:update:end}
		\State \textbf{return} $\asubtensormax$
		
	\end{algorithmic}
\end{algorithm}

\textbf{(b) Reorder (Lines~\ref{alg:method:increment:line:reorder:start}-\ref{alg:method:increment:line:reorder:end})}: The goal of this step is to reorder the slice indices located in the region $[j_{L}, j_{H}]$  so that $\pi$ remains as a \order in $\TT$ after the change $\increment$.
Let $\TT'$ be the updated $\TT$ and $\pi'$ be the updated $\pi$ to distinguish them with $\TT$ and $\pi$ before the update.
We get $\pi'$ from $\pi$ by reordering the slice indices in
\begin{equation} 
R=\{\amem \in \aset: \piinvfun{\amem} \in [j_{L}, j_{H}]\} \label{eq:increment:R}
\end{equation}
so that the following condition is met for every $j \in [j_{L},j_{H}]$ and the corresponding $q=\pifunnew{j}$:
\begin{equation}
\hspace{-2mm} \dfun{\TT'(\aset_{\pi',q}),q} =\min\nolimits_{\amemtwo \in R \cap \aset_{\pi',q}} \dfun{\TT'(\aset_{\pi',q}),\amemtwo}. \label{eq:increment:range}
\end{equation}
This guarantees that $\pi'$ is a \order in $\TT'$, as shown in Lemma~\ref{lemma:increment:order}.
\begin{lemma} \label{lemma:increment:order}
	Let $\pi$ be a \order in $\TT$, and let $\TT'$ be $\TT$ after a change $\increment$. 
	For $R$ $(Eq.~\eqref{eq:increment:R})$ defined on $j_{L}$ and $j_{H}$ , $(Eq.~\eqref{eq:increment:min}$ and $Eq.~\eqref{eq:increment:max})$,
	let $\pi'$ be an ordering of slice indices $Q$ where $\forall j\notin[j_{L},j_{H}],$ $\pi'(j)=\pi(j)$ and $\forall j\in[j_{L},j_{H}]$, Eq.~\eqref{eq:increment:range} holds.
	Then, $\pi'$ is a \order in $\TT'$.
	\vspace{0.5mm} \\
	\noindent {\it Proof.} See Section~B of the supplementary document \cite{supple}. \qed
\end{lemma}


\textbf{(c) Update-Subtensor (Lines~\ref{alg:method:increment:line:update:start}-\ref{alg:method:increment:line:update:end})}: In this step, we update the maintained dense subtensor $\asubtensormax$ when two conditions are met.
We first check $\cmax\geq \density{\asubtensormax}$, which takes $O(1)$ if we maintain $\density{\asubtensormax}$, since $\cmax < \density{\asubtensormax}$ entails that the updated entry $\avalue$ is not in the densest subtensor (see the proof of Theorem~\ref{thm:increment:accuracy} for details).
We then check if there are changes in $S_{max}$, obtained by \textsc{find-Slices()}.
This takes only $O(|Q|)$, as shown in Theorem~\ref{thm:static:time}. 
Even if both conditions are met, updating $\asubtensormax$ is simply to construct $\asubtensormax$ from given $\asubsetmax$ instead of finding $\asubtensormax$ from scratch.
This conditional update reduces computation but still preserves Property~\ref{property:dynamic}, as formalized in Lemma~\ref{lemma:increment:densest} and Theorem~\ref{thm:increment:accuracy}.

\begin{lemma} \label{lemma:increment:densest}
	Consider a \order $\pi$ in $\TT$. For every entry $\avalue$ with index $\aindex$ belonging to the densest subtensor, $\forall n\in [N]$, $\cpifun{(n,i_{n})}\geq \densityopt$ holds.
	\vspace{0.5mm} \\
	\noindent {\it Proof.} Let $\asubtensoropt$ be a subtensor with the maximum density, i.e., $\density{\asubtensoropt}=\densityopt$.
	Let $\amem^{*}\in \asubsetopt$ be satisfying
	that $\forall \amem\in \asubsetopt$, $\pi^{-1}(\amem^{*})\leq\piinvfun{\amem}$. 
	For any entry $\avalue$ in $\asubtensoropt$ with index $\aindex$ and any $\amem\in \{(n,i_{n}):  n \in [N]\}$,
	our goal is to show $c_{\pi}(\amem)\geq\density{\asubtensoropt}$, which we show
	as $c_{\pi}(\amem)\geq d_{\pi}(\amem^{*})\geq d(\asubtensoropt,\amem^{*})$ $\geq\density{\asubtensoropt}$.
	
	First, $c_{\pi}(\amem)\geq d_{\pi}(\amem^{*})$ is from the definition
	of $c_{\pi}(\amem)$ and $\pi^{-1}(\amem^{*})\leq\piinvfun{\amem}$. Second,
	from $\asubsetopt\subset \aset_{\pi,\amem^{*}}$,  $d_{\pi}(\amem^{*})=d(\TT(\aset_{\pi,\amem^{*}}),$ $\amem^{*})\geq d(\asubtensoropt,\amem^{*})$ holds.
	Third, $d(\asubtensoropt,\amem^{*})\geq\density{\asubtensoropt}$ is from Lemma~\ref{lemma:maxdensity}.
	From these, $c_{\pi}(\amem)\geq\density{\asubtensoropt}$ holds. \qed
\end{lemma}

\begin{theorem}[Accuracy Guarantee of Algorithm~\ref{alg:method:increment}] \label{thm:increment:accuracy}
Algorithm~\ref{alg:method:increment} preserves Property~\ref{property:dynamic}, and thus $\density{\asubtensormax}\geq \densityopt/N$ holds after Algorithm~\ref{alg:method:increment} terminates.
\vspace{0.5mm} \\
\noindent {\it Proof.} We assume that Property~\ref{property:dynamic} holds and prove that it still holds after Algorithm~\ref{alg:method:increment} is executed.
First, the ordering $\pi$ remains to be a \order in $\TT$ by Lemma~\ref{lemma:increment:order}.
Second, we show $\density{\asubtensormax}$ $\geq \densityopt/N$.
If the condition in line~\ref{alg:method:increment:line:cond} of Algorithm~\ref{alg:method:increment} is met, $\asubtensormax$ is set to the subtensor with the maximum density in $\{\asubtensor:\amem \in \aset\}$ by \textsc{Find-Slices()}. By Lemma~\ref{lemma:static:order}, $\density{\asubtensormax}\geq \densityopt/N$.
If the condition in line~\ref{alg:method:increment:line:cond} is not met, for the changed entry $\avalue$ with index $\aindex$, by the definition of $j_{L}$, there exists $n\in[N]$ such that $\pi(j_{L})=(n, i_{n})$.
Since $j_{L} \leq j_{H}$, $c_{\pi}((n, i_{n}))=\cpifun{\pifun{j_{L}}}\leq \cpifun{\pifun{j_{H}}}  = c_{max} < \density{\asubtensormax}\leq\densityopt$.
Then, by Lemma~\ref{lemma:increment:densest}, $\avalue$ does not belong to the densest subtensor, which thus remains the same after the change $\increment$.
Since $\density{\asubtensormax}$ never decreases,
$\density{\asubtensormax}$ $\geq\densityopt/N$ still holds by Property~\ref{property:dynamic}, which we assume.
Property~\ref{property:dynamic} is preserved because its two conditions are met. \qed
\end{theorem}

Theorem~\ref{thm:method:time} gives the time complexity of Algorithm~\ref{alg:method:increment}.
In the worst case (i.e., $R=\aset$), this becomes $O(|\aset|\log|\aset|$ $+\annz{\TT}N)$, which is the time complexity of Algorithm~\ref{alg:static}.
In practice, however, $R$ is much smaller than $Q$, and updating $\asubtensormax$ happens rarely. Thus, in our experiments, Algorithm~\ref{alg:method:increment} scaled sub-linearly with $\annz{\TT}$ (see Section~\ref{sec:exp:scalability}). 
\begin{theorem}[Time Complexity of Algorithms~\ref{alg:method:increment} and \ref{alg:method:decrement}] \label{thm:method:time} 
	Let $\TT(\amem)$ be the set of the non-zero entries in the slice with index $q$ in $\TT$.
	The time complexity of Algorithms~\ref{alg:method:increment} and \ref{alg:method:decrement} is $O(|R|\log|R|+|\aset|+N|\bigcup_{\amem\in R}\TT(\amem)|+N|\bigcup_{\amem\in \asubsetmax}\TT(\amem)|)$.
\vspace{0.5mm} \\
\noindent {\it Proof.}
	Assume that, for each slice, the list of the non-zero entries in the slice is stored, and let $\amem_{f}=\pi(j_{L})$.
	Computing $j_{L}$, $j_{H}$, and $R$ takes $O(|R|)$.
	Assume we use a Fibonacci heap to find slices with minimum slice sum (line~\ref{alg:increment:line:minslice} of Algorithm~\ref{alg:method:increment}).
	Computing the slice sum of every slice in $R$  in $\TT(\aset_{\pi,\amem_{f}})$ takes $O(N|\bigcup_{\amem\in R}\TT(\amem)|)$.
	Then, constructing a Fibonacci heap where each value is a slice index in $R$ and the corresponding key is the slice sum of the slice in $\TT(\aset_{\pi,\amem_{f}})$  takes $O(|R|)$.
	Popping the index of a slice with minimum slice sum, which takes $O(\log|R|)$, happens $|R|$ times, and thus we get $O(|R|\log|R|)$.
	Whenever a slice index is popped we have to update the slice sums of its dependent slices in $R$ (two slices are dependent if they have common non-zero entries).	
	Updating the slice sum of each dependent slice, which takes $O(1)$ in a Fibonacci heap, happens at most $O(N|\bigcup_{\amem\in R}\TT(\amem)|)$ times, and thus we get $O(N|\bigcup_{\amem\in R}\TT(\amem)|)$.	
	On the other hand,  by Lemma~\ref{lemma:time:find}, \textsc{Find-Slices()} and constructing $\asubtensormax$ from $\asubsetmax$ take $O(|\aset|+N|\bigcup_{\amem\in \asubsetmax}\TT(\amem)|).$
	Hence, the time complexity of Algorithms~2 and 3 is the sum of all the costs, which is $O(|R|\log|R|+|\aset|+N|\bigcup_{\amem\in R}\TT(\amem)|+N|\bigcup_{\amem\in \asubsetmax}\TT(\amem)|)$. \qed
\end{theorem}

\subsubsection{Decrement of Entry Values.}
\label{sec:method:dynamic:dec}
As in the previous section, assume that a tensor $\TT$, a \order $\pi$ (also $\dfunnoarg$ and $\cfunnoarg$), and a dense subtensor $\asubtensormax$ satisfying Property~\ref{property:dynamic} are maintained.
(such $\pi$, $\dfunnoarg$, $\cfunnoarg$, and $\asubtensormax$ can be initialized by Algorithm~\ref{alg:static} if we start from scratch).
Algorithm~\ref{alg:method:decrement} describes
the response of \methodS to $\decrement$, a decrement of entry $\avalue$ by $\delta > 0$, for satisfying Property~\ref{property:dynamic}.
Algorithm~\ref{alg:method:decrement} has the same structure of Algorithm~\ref{alg:method:increment}, while they are different in the reordered region of $\pi$ and the conditions for updating the dense subtensor.
The differences are explained below.

For a change $\decrement$, we find the region $[j_{L}, j_{H}]$ of the domain of $\pi$ that may need to be reordered.
Let  $C=\{(n,i_{n}): n\in[N]\}$ be the indices of the slices composing the changed entry $\avalue$, and let $\amem_{f}=\argmin_{\amem\in C}\piinvfun{\amem}$ be the one located first in $\pi$ among $C$.
Then, let $M_{min}=\{\amem \in \aset : \adegree > \cpifun{\amem_{f}}-\delta \}$ and $M_{max}=\{\amem \in \aset :\piinvfun{\amem} > \piinvfun{\amem_{f}}, \ \adegree \geq \cpifun{\amem_{f}} \}$.
Note that $M_{min}\neq \emptyset$ since, by the definition of $c_{\pi}(\cdot)$, there exists $\amem \in \aset$ where $\piinvfun{\amem}\leq \piinvfun{\amem_{f}}$ and $\adegree=c_{\pi}(\amem_{f})$.
Then, $j_{L}$ and $j_{H}$ are:
\begin{align}
	j_{L}= &  \min\nolimits_{\amem \in M_{min}} \piinvfun{\amem}, \label{eq:decrement:min}\\ 
	j_{H}= & \begin{cases}
		\min_{\amem \in M_{max}} \piinvfun{\amem}-1 & \text{if}\ M_{max}\neq \emptyset, \\
		|\aset|  \text{ (i.e., the last index)} & \text{otherwise.} \label{eq:decrement:max}
	\end{cases}
\end{align}

As in the increment case, we update $\pi$, to remain it as a \order, by reordering the slice indices located in $[j_{L},j_{H}]$ of $\pi$.
Let $\TT'$ be the updated $\TT$ and $\pi'$ be the updated $\pi$ to distinguish them with $\TT$ and $\pi$ before the update.
Only the slice indices in $R=\{\amem \in \aset: \piinvfun{\amem} \in [j_{L}, j_{H}]\}$ are reordered in $\pi$ so that Eq.~\eqref{eq:increment:range} is met for every $j \in [j_{L},j_{H}]$.
This guarantees that $\pi'$ is a \order, as formalized in Lemma~\ref{lemma:decrement:order}.

\begin{algorithm}[t]
	\small
	\caption{\methodS in the case of decrement}\label{alg:method:decrement}
	\begin{algorithmic}[1] 
		\Require (1) current tensor: $\TT$ with slice indices $Q$
		\Statex \ \ \quad (2) current dense subtensor: $\asubtensormax$ with Property~\ref{property:dynamic}
		\Statex \ \ \quad (3) current \order: $\pifunnoarg$ (also $\cfunnoarg$ and $\dfunnoarg$)
		\Statex \ \ \quad (4) a change in $\TT$: $\decrement$
		\Ensure updated dense subtensor$\asubtensormax$
		\State $\avalue \gets \avalue - \delta$ 
		\State compute $j_{L}$ and $j_{H}$ by Eq. \eqref{eq:decrement:min} and \eqref{eq:decrement:max} \hfill \bluecolor{$\vartriangleright$  \textsc{[Find-Region]}}
		\State Lines~\ref{alg:method:increment:line:reorder:start}-\ref{alg:method:increment:line:reorder:end} of Algorithm~\ref{alg:method:increment}\hfill \bluecolor{$\vartriangleright$  \textsc{[Reorder]}}
		\If{$\avalue$ is in $\asubtensormax$} \label{alg:method:decrement:line:cond}\hfill \bluecolor{$\vartriangleright$  \textsc{[Update-Subtensor]}}
		\State $S'\gets$\Call{Find-Slices}{}() in Algorithm~\ref{alg:static} \hfill \bluecolor{$\vartriangleright$ time complexity: $O(|Q|)$} 
		\If{$\asubsetmax \neq S'$}{ $\asubtensormax\gets\TT(S')$} \EndIf
		\EndIf 
		\State \textbf{return} $\asubtensormax$
	\end{algorithmic}
\end{algorithm}

\begin{lemma} \label{lemma:decrement:order}
	Let $\pi$ be a \order in $\TT$, and let $\TT'$ be $\TT$ after a change $\decrement$. 
	For $R$ $(Eq.~\eqref{eq:increment:R})$ defined on $j_{L}$ and $j_{H}$ $(Eq.~\eqref{eq:decrement:min}$ and $Eq.~\eqref{eq:decrement:max})$,
	let $\pi'$ be an ordering of slice indices $Q$ where $\forall j\notin[j_{L},j_{H}],$ $\pifunnew{j}=\pifun{j}$ and $\forall j\in[j_{L},j_{H}]$, Eq.~\eqref{eq:increment:range} holds.
	Then, $\pi'$ is a \order in $\TT'$.
\vspace{0.5mm} \\
\noindent {\it Proof.}
	See Section~B of the supplementary document \cite{supple}. \qed
\end{lemma}

The last step of Algorithm~\ref{alg:method:decrement} is to conditionally and rapidly update the maintained dense subtensor $\asubtensormax$ using $\pi$. 
The subtensor $\asubtensormax$ is updated if entry $\avalue$ belongs to $\asubtensormax$ (i.e., if $\density{\asubtensormax}$ decreases by the change $\decrement$) and there are changes in $S_{max}$, obtained by \textsc{find-Slices()}.
Checking these conditions takes only $O(|Q|)$, as in the increment case.
Even if $\asubtensormax$ is updated, it is just constructing $\asubtensormax$ from given $\asubsetmax$, instead of finding $\asubtensormax$ from scratch.

Algorithm~\ref{alg:method:decrement} preserves Property~\ref{property:dynamic}, as shown in Theorem~\ref{thm:decrement:accuracy}, and has the same time complexity of Algorithm~\ref{alg:method:increment} in Theorem~\ref{thm:method:time}.

\begin{theorem}[Accuracy Guarantee of Algorithm~\ref{alg:method:decrement}] \label{thm:decrement:accuracy}
	Algorithm~\ref{alg:method:decrement} preserves Property~\ref{property:dynamic}.
	Thus, $\density{\asubtensormax}\geq \densityopt/N$ holds after Algorithm~\ref{alg:method:decrement} terminates.
\vspace{0.5mm} \\
\noindent {\it Proof.}
	We assume that Property~\ref{property:dynamic} holds and prove that it still holds after Algorithm~\ref{alg:method:decrement} is executed.
	First, the ordering $\pi$ remains to be a \order in $\TT$ by Lemma~\ref{lemma:decrement:order}.
	Second, we show  $\density{\asubtensormax} \geq \densityopt/N$.
	If the condition in line~\ref{alg:method:decrement:line:cond} of Algorithm~\ref{alg:method:decrement} is met, $\asubtensormax$ is set to the subtensor with the maximum density in $\{\asubtensor:\amem \in \aset\}$ by \textsc{Find-Slices()}. By Lemma~\ref{lemma:static:order}, $\density{\asubtensormax}\geq \densityopt/N$.
	If the condition is not met, $\density{\asubtensormax}$ remains the same, while $\densityopt$ never increases. Hence,
	$\density{\asubtensormax} \geq \densityopt/N$ still holds by Property~\ref{property:dynamic}, which we assume.
	Since its two conditions are met, Property~\ref{property:dynamic} is preserved. \qed
\end{theorem}

\subsubsection{Increase or Decrease of Size.}
\label{sec:method:dynamic:size}
\methodS also supports the increase and decrease of the size of the input tensor.
The increase of the
size of $\TT$ corresponds to the addition of new slices to $\TT$.
For example, if the length of the $n$th mode of $\TT$ increases from $I_{n}$ to $I_{n}+1$, the index $q=(n, I_{n}+1)$ of the new slice is added to $Q$ and the first position of $\pi$.
We also set $\adegree$ and $\acore$ to $0$.
Then, if there exist non-zero entries in the new slice, they are handled one by one by Algorithm~\ref{alg:method:increment}. 
Likewise, when size decreases, we first handle the removed non-zero entries one by one by Algorithm~\ref{alg:method:decrement}. Then, we remove the indices of the removed slices from $Q$ and $\pi$.

\subsection{\textsc{\Large \methodA}: Suddenly Emerging Dense-Subtensor Detection}
\label{sec:method:alert}

\begin{figure}[t]
	\centering
	\includegraphics[width=\linewidth]{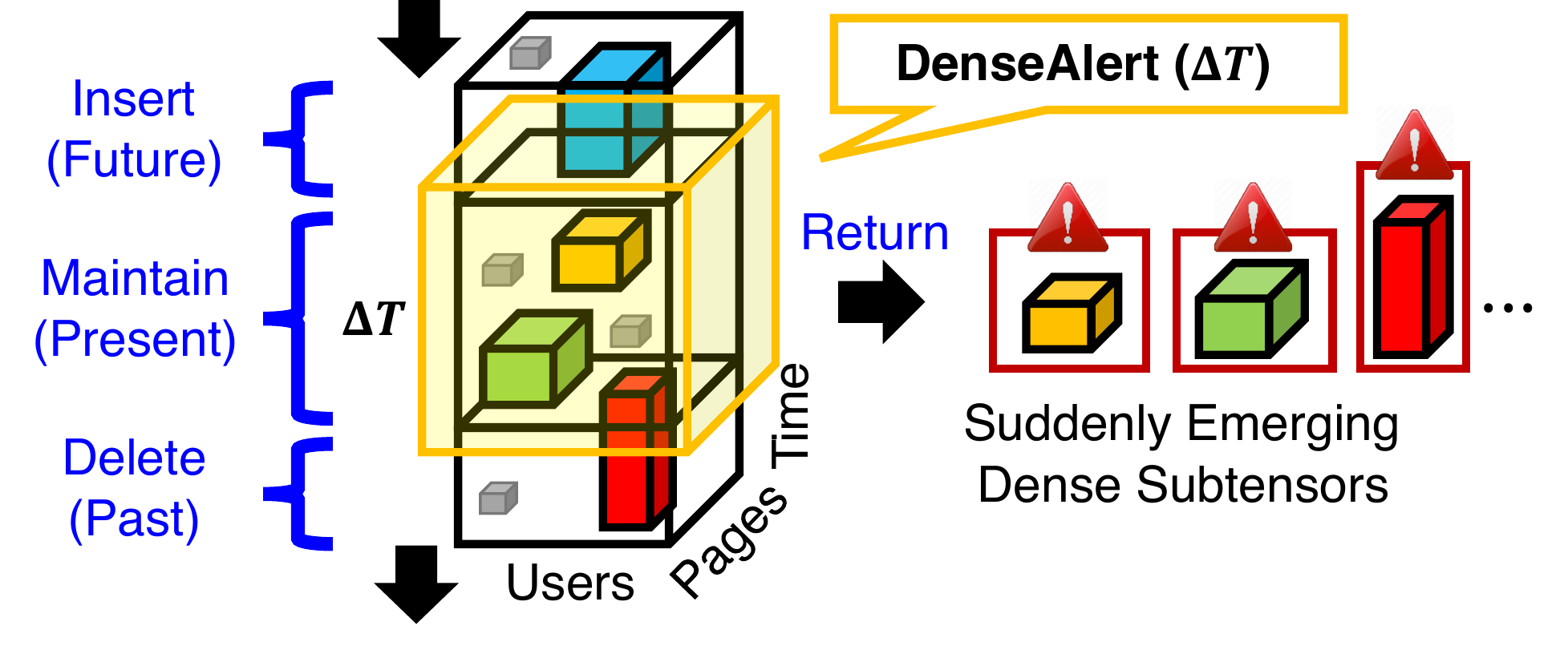} 
	\caption{\label{fig:method}
		\methodA with Wikipedia Revision History (Example~\ref{example}). $\methodA$ (yellow box in the figure) spots dense subtensors formed within $\Delta T$ time units.
	}
\end{figure}

Based on \methodS, we propose \methodA, an incremental algorithm for detecting suddenly emerging dense subtensors.
For a stream of increments in the input tensor $\TT$,
\methodA maintains $\TT_{\Delta T}$, a tensor where the value of each entry is the increment of the value of the corresponding entry in $\TT$ in last $\Delta T$ time units (see Problem~\ref{problem:time} in Section~\ref{sec:prelim:problem}), as described in Figure~\ref{fig:method} and Algorithm~\ref{alg:method:realtime},
To maintain $\TT_{\Delta T}$ and a dense subtensor in it, \methodA applies increments by \methodS (line~\ref{alg:method:real-time:line:increment}), and undoes the increments after $\Delta T$ time units also by \methodS (lines~\ref{alg:method:real-time:line:schedule} and \ref{alg:method:real-time:line:decrement}).
The accuracy of \methodA, formalized in Theorem~\ref{thm:realtime:accuracy}, is implied from the accuracy of \methodS. 

\begin{algorithm}[t]
	\small
	\caption{\methodA for sudden dense subtensors}\label{alg:method:realtime}
	\begin{algorithmic}[1] 
		\Require (1) sequence of increments in $\TT$
		\Statex \ \ \quad (2) time window: $\Delta T$
		\Ensure suddenly emerging dense subtensors
		\State run Algorithm~\ref{alg:static} with a zero tensor
		\State wait until the next change happens at time $T$
		\If{the change is $\increment$}
		\State run \methodS (Algorithm~\ref{alg:method:increment}) \label{alg:method:real-time:line:increment}
		\State schedule $\decrement$ at time $T+\Delta T$ \label{alg:method:real-time:line:schedule}
		\ElsIf{the change is $\decrement$}
		\State run \methodS (Algorithm~\ref{alg:method:decrement}) \label{alg:method:real-time:line:decrement}
		\EndIf
		\State report the current dense subtensor
		\State {\bf goto} Line~2
	\end{algorithmic}
\end{algorithm}

\begin{theorem}[Accuracy Guarantee of Algorithm~\ref{alg:method:realtime}] \label{thm:realtime:accuracy} 
	Let $\Delta\densityopt$ be the density of the densest subtensor in the $N$-way tensor $\TT_{\Delta T}$.
	The subtensor maintained by
	Algorithm~\ref{alg:method:realtime} has density at least $\Delta\densityopt/N$.
\vspace{0.5mm} \\
\noindent {\it Proof.}
	By Theorems~\ref{thm:increment:accuracy} and \ref{thm:decrement:accuracy}, \methodA, which uses \methodS for updates, maintains a subtensor with density at least $1/N$ of the density of the densest subtensor. \qed
\end{theorem}
The time complexity of \methodA is also obtained from Theorem~\ref{thm:method:time} by simply replacing $\TT$ with $\TT_{\Delta T}$.
\methodA needs to store only $\TT_{\Delta T}$ (i.e., the changes in the last $\Delta T$ units) in memory at a time. \methodA discards older changes.

%% file: 040experiments.tex
\begin{table}[t]
	\centering
	\small
	\caption{\label{tab:data:real} Summary of real-world tensor datasets.}
	\begin{tabular}{l|l|c|c}
		\toprule
		{\bf Name} & {\bf Size} & $\mathbf{|Q|}$ &  $\mathbf{\annz{\TT}}$ \\
		\midrule
		\multicolumn{4}{l}{Ratings: users $\times$ items $\times$ timestamps $\times$ ratings $\rightarrow$ \#reviews} \\
		\midrule
		Yelp \cite{datayelpchallenge} & 552K $\times$ 77.1K $\times$ 3.80K $\times$ 5 & 633K & 2.23M \\
		Android \cite{mcauley2015inferring} & 1.32M $\times$ 61.3K $\times$ 1.28K $\times$ 5 & 1.39M & 2.64M \\
		YahooM. \cite{dror2012yahoo} & 1.00M $\times$ 625K $\times$ 84.4K $\times$ 101& 1.71M &  253M \\
		\midrule
		\multicolumn{4}{l}{Wikipedia edit history: users $\times$ pages $\times$ timestamps $\rightarrow$ \#edits}\\
		\midrule
		KoWiki \cite{shin2016mzoom} & 470K $\times$ 1.18M $\times$ 101K & 1.80M & 11.0M \\
		EnWiki \cite{shin2016mzoom}  & 44.1M $\times$ 38.5M $\times$  129K & 82.8M & 483M\\
		\midrule
		\multicolumn{4}{l}{Social networks: users $\times$ users $\times$ timestamps $\rightarrow$ \#interactions}\\
		\midrule
		Youtube \cite{mislove-2007-socialnetworks} & 3.22M $\times$ 3.22M $\times$ 203 & 6.45M & 18.7M \\
		SMS & 1.25M $\times$ 7.00M $\times$ 4.39K & 8.26M & 103M \\
		\midrule
		\multicolumn{4}{l}{{TCP dumps: IPs $\times$ IPs $\times$ timestamps $\rightarrow$ \#connections}}\\
		\midrule
		TCP \cite{lippmann2000evaluating} & 9.48K $\times$ 23.4K $\times$ 46.6K & 79.5K &  522K \\
		\bottomrule
	\end{tabular}
\end{table}

We design experiments to answer the following questions:
\bit
	\item \textbf{Q1. Speed}: How fast are updates in \methodS compared to batch algorithms?
	\item \textbf{Q2. Accuracy}: How accurately does \methodS maintain a dense subtensor?
	\item \textbf{Q3. Scalability}: How does the running time of \methodS increase as input tensors grow?
	\item \textbf{Q4. Effectiveness}: Which anomalies or fraudsters does \methodA spot in real-world tensors?
\eit

\subsection{Experimental Settings.}
{\bf Machine:} We ran all experiments on a machine with 2.67GHz Intel Xeon E7-8837 CPUs and 1TB memory (up to 85GB was used by our algorithms).

\noindent{\bf Data:} Table~\ref{tab:data:real} lists the real-world tensors used in our experiments. {\it Ratings} are
4-way tensors (users, items, timestamps, ratings) where entry values are the number of reviews. {\it Wikipedia edit history} is 3-way tensors (users, pages, timestamps) where entry values are the number of edits.
{\it Social networks} are 3-way tensors (source users, destination users, timestamps) where entry values are the number of interactions.
{\it TCP dumps} are 3-way tensors (source IPs, destination IPs, timestamps) where entry values are the number of TCP connections.
Timestamps are in dates in Yelp and Youtube, in minutes in TCP, and in hours in the others.

\begin{figure*}[t]
	\centering
	\includegraphics[width=0.625\linewidth]{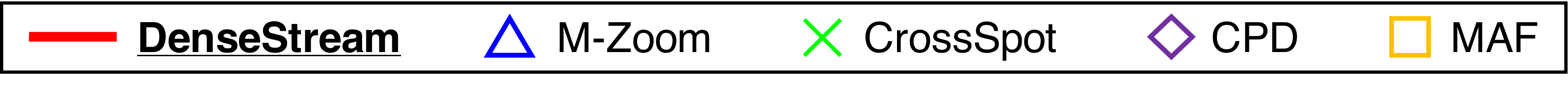} \\
	\vspace{-1.5mm}
	\subfigure[SMS (part)]{
		\includegraphics[width=0.21\linewidth]{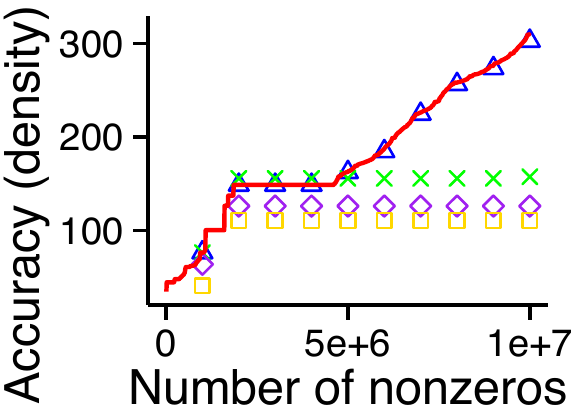}
	}
	\subfigure[Youtube]{
		\includegraphics[width=0.21\linewidth]{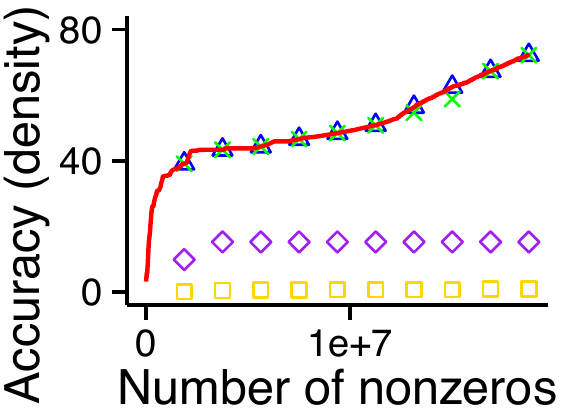}
	}
	\subfigure[EnWiki (part)]{
		\includegraphics[width=0.21\linewidth]{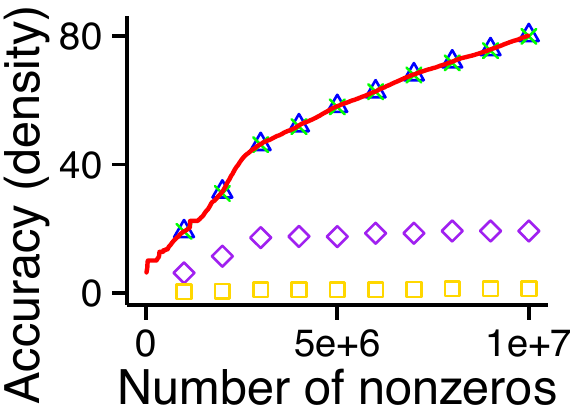}
	}
	\subfigure[KoWiki]{
		\includegraphics[width=0.21\linewidth]{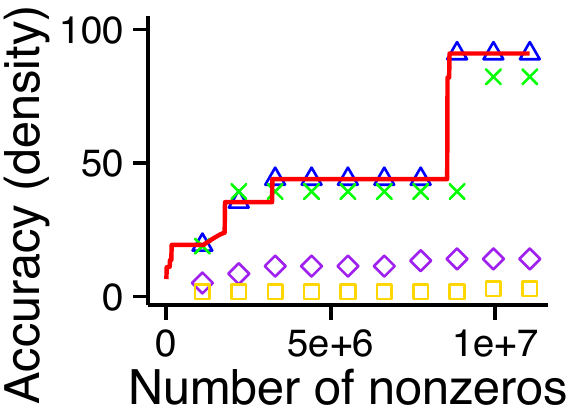}
	} \\
	\vspace{-3.5mm}
	\subfigure[YahooM. (part)]{
		\includegraphics[width=0.21\linewidth]{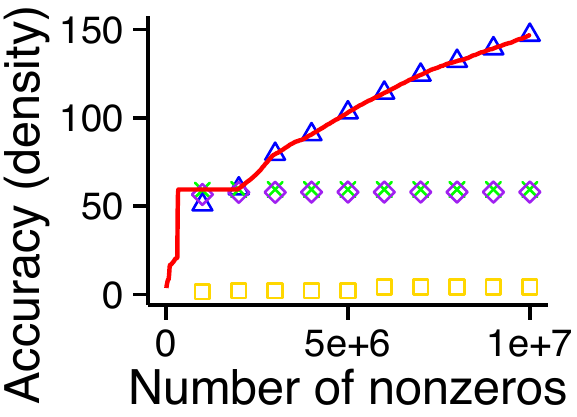}
	}
	\subfigure[Android]{
		\includegraphics[width=0.21\linewidth]{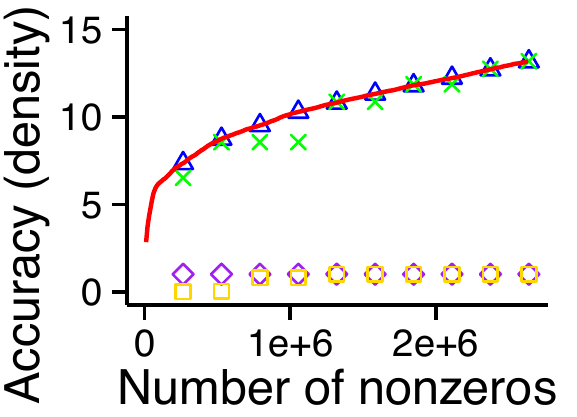}
	}
	\subfigure[Yelp]{
		\includegraphics[width=0.21\linewidth]{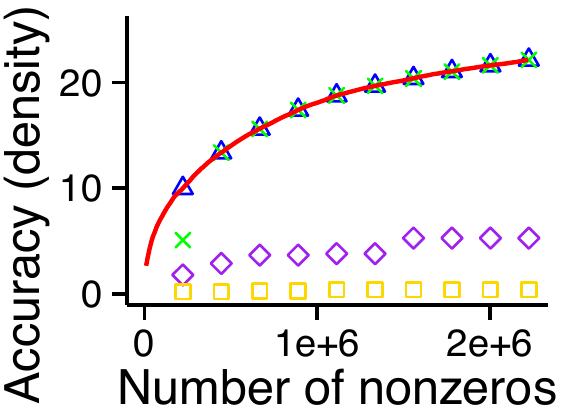}
	}
	\subfigure[TCP]{
		\includegraphics[width=0.21\linewidth]{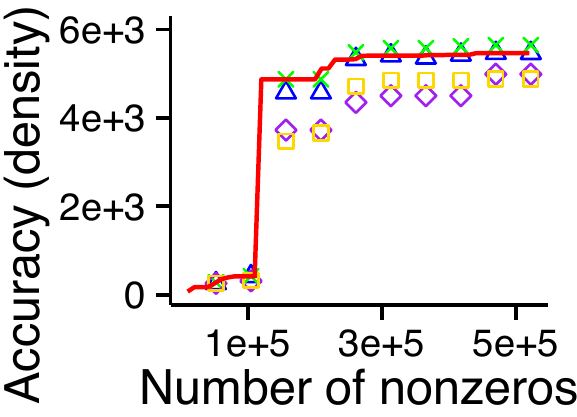}
	}
	\\
	\vspace{-2mm}
	\caption{\label{fig:accuracy}
		\underline{\smash{ \methodS is `any-time' and accurate.}}
		While tensors grow, \methodS maintains and instantly updates a dense subtensor, whereas
		batch algorithms update dense subtensors only once in a time interval.
		Subtensors maintained by \methodS have density (red lines) similar to the density (points) of the subtensors found by the best batch algorithms.
	}
\end{figure*}

\noindent{\bf Implementations:} We implemented dense-subtensor detection algorithms for comparison.
We implemented our algorithms, \mzoom \cite{shin2016mzoom}, and \cross \cite{jiang2015general} in Java, while we used Tensor
Toolbox \cite{TTB_Software} for CP decomposition (CPD)\footnote{
	Let $\MA^{(1)}\in \mathbb{R}^{I_{1}\times k}$, $\MA^{(2)}\in \mathbb{R}^{I_{2}\times k}$, ..., $\MA^{(N)}\in \mathbb{R}^{I_{N}\times k}$ be the factor matrices obtained by the rank-$k$ CP Decomposition of $\TR$.
	For each $j\in [k]$, we form a subtensor with every slice with index $(n,i_{n})$ where the $(i_{n},j)$-th entry of $\MA^{(n)}$ is at least $1/\sqrt{I_{n}}$.} and
MAF \cite{maruhashi2011multiaspectforensics}.
In all the implementations, a sparse tensor format was used so that the space usage is proportional to the number of non-zero entries.
As in \cite{shin2016mzoom},
we used a variant of \cross which maximizes the density measure defined in Definition~\ref{defn:density} and uses CPD for seed selection.
For each batch algorithm, we reported the densest one after finding three dense subtensors.

\begin{figure}[t]
	\centering
	\vspace{1mm}
	\includegraphics[width=\linewidth]{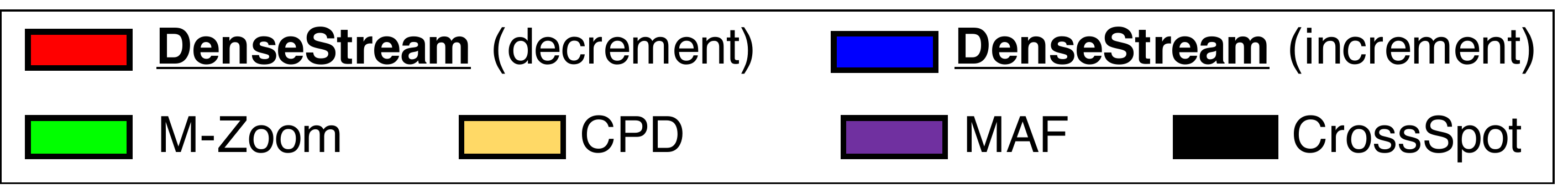} \\
	\includegraphics[width=\linewidth]{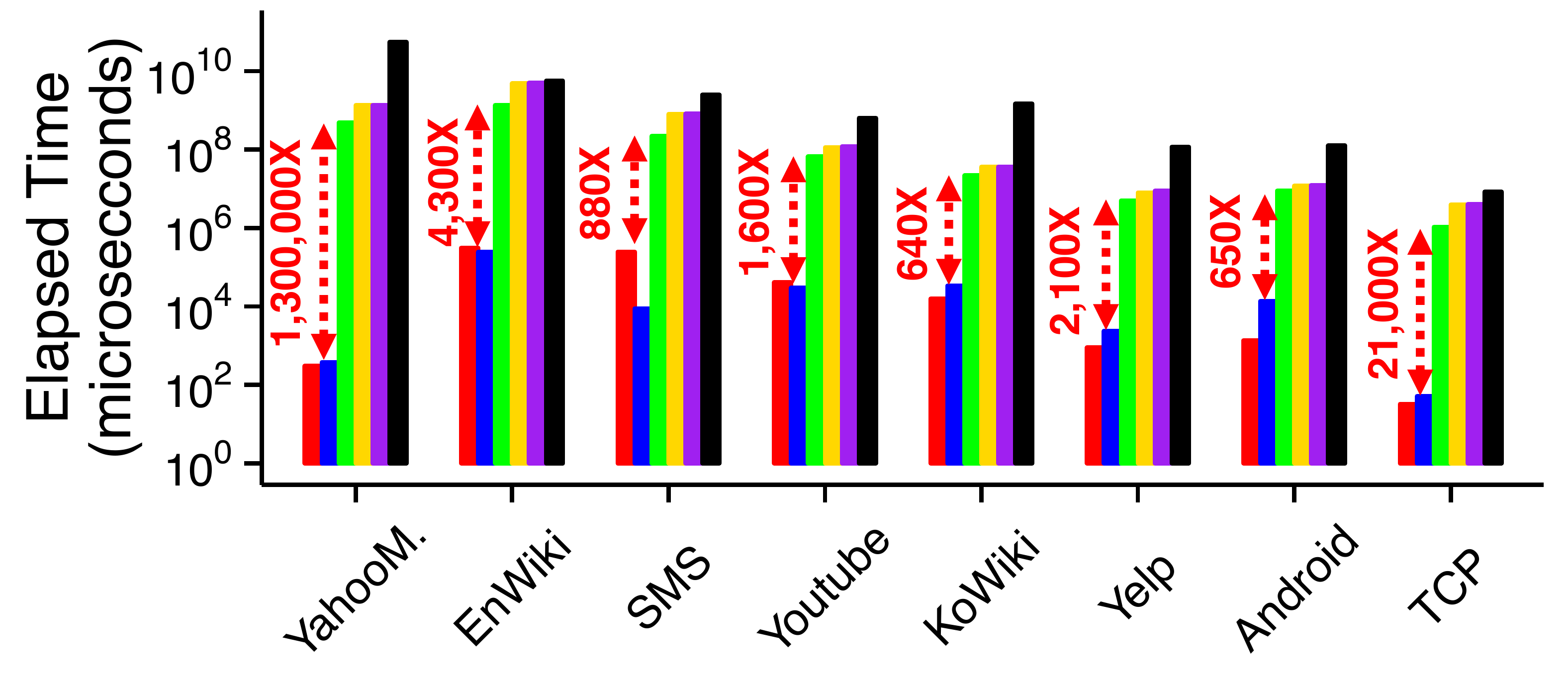} \\
	\vspace{-2mm}
	\caption{\label{fig:speed}
		\underline{\smash{\methodS outperforms batch algorithms.}}
		An update in \methodS was up to {\bf a million times faster} than the fastest batch algorithm.
	}
\end{figure}

\subsection{Q1. Speed} \label{sec:exp:speed}
We show that updating a dense subtensor by \methodS is significantly faster than running batch algorithms from scratch.
For each tensor stream, we averaged the update times for processing the last 10,000 changes corresponding to increments  (blue bars in Figure~\ref{fig:speed}).
Likewise, we averaged the update times for undoing the first 10,000 increments, i.e., decreasing the values of the oldest entries (red bars in Figure~\ref{fig:speed}).
Then, we compared them to the time taken for running batch algorithms on the final tensor that each tensor stream results in.
As seen in Figure~\ref{fig:speed}, updates in \methodS were up to a million times faster than the fastest batch algorithm.
The speed-up was particularly high in sparse tensors having a widespread slice sum distribution (thus having a small reordered region $R$), as we can expect from Theorem~\ref{thm:method:time}.

On the other hand, the update time in \methodA, which uses \methodS as a sub-procedure, was similar to that in \methodS when the time interval $\Delta T = \infty$, and was less with smaller $\Delta T$.
This is since the average number of non-zero entries maintained is proportional to  $\Delta T$. 

\subsection{Q2. Accuracy}
\label{sec:exp:accuracy}
This experiment demonstrates the accuracy of \methodS. From this, the accuracy of \methodA, which uses \methodS as a sub-procedure, is also obtained.
We tracked the density of the dense subtensor maintained by \methodS while each tensor grows, and compared it to the densities of the dense subtensors found by batch algorithms.
As seen in Figure~\ref{fig:accuracy}, the subtensors that \methodS maintained had density (red lines) similar to the density (points) of the subtensors found by the best batch algorithms.
Moreover, \methodS is `any time'; that is, as seen in Figure~\ref{fig:accuracy}, \methodS updates the dense subtensor instantly, while the batch algorithms cannot update their dense subtensors until the next batch processes end.
\methodS also maintains a dense subtensor accurately when the values of entries decrease, as shown in Section~C of the supplementary document \cite{supple}.

The accuracy and speed (measured in Section~\ref{sec:exp:speed}) of the algorithms in Yelp Dataset are shown in Figure~\ref{fig:crown:tradeoff} in Section~\ref{sec:intro}.
\methodS significantly reduces the time gap between the emergence and the detection of a dense subtensor, without losing accuracy.


\subsection{Q3. Scalability} 
\label{sec:exp:scalability}
We demonstrate the high scalability of \methodS by measuring how rapidly its update time increases as a tensor grows.
For this experiment,
we used a $10^5\times10^5\times10^5$ random tensor stream that has a realistic power-law slice sum distribution in each mode.
As seen in Figure~\ref{fig:crown:scalable} in Section~\ref{sec:intro}, update times, for both types of changes, scaled sub-linearly with the number of nonzero entries. 
Note that \methodA, which uses \methodS as a sub-procedure, has the same scalability.

\subsection{Q4. Effectiveness}
\label{sec:exp:effective}
In this section, we show the effectiveness of \methodA in practice. We focus on \methodA, which spots suddenly emerging dense subtensors overlooked by existing methods, rather than \methodS, which is much faster but eventually finds a similar subtensor with previous algorithms, especially \cite{shin2016mzoom}.

\begin{figure}[t]
	\centering
	\includegraphics[width=0.8\linewidth]{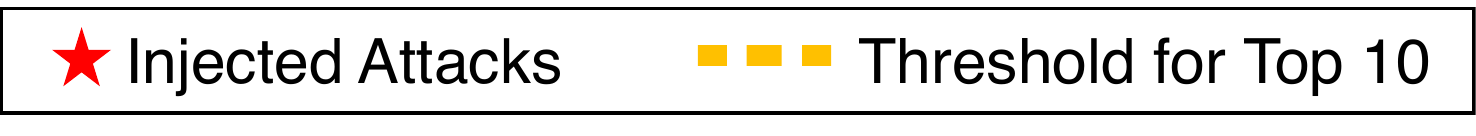} \\
	\vspace{-1.5mm}
	\subfigure[\methodA  in Yelp (left) and Android (right)]{
		\includegraphics[width=0.45\linewidth]{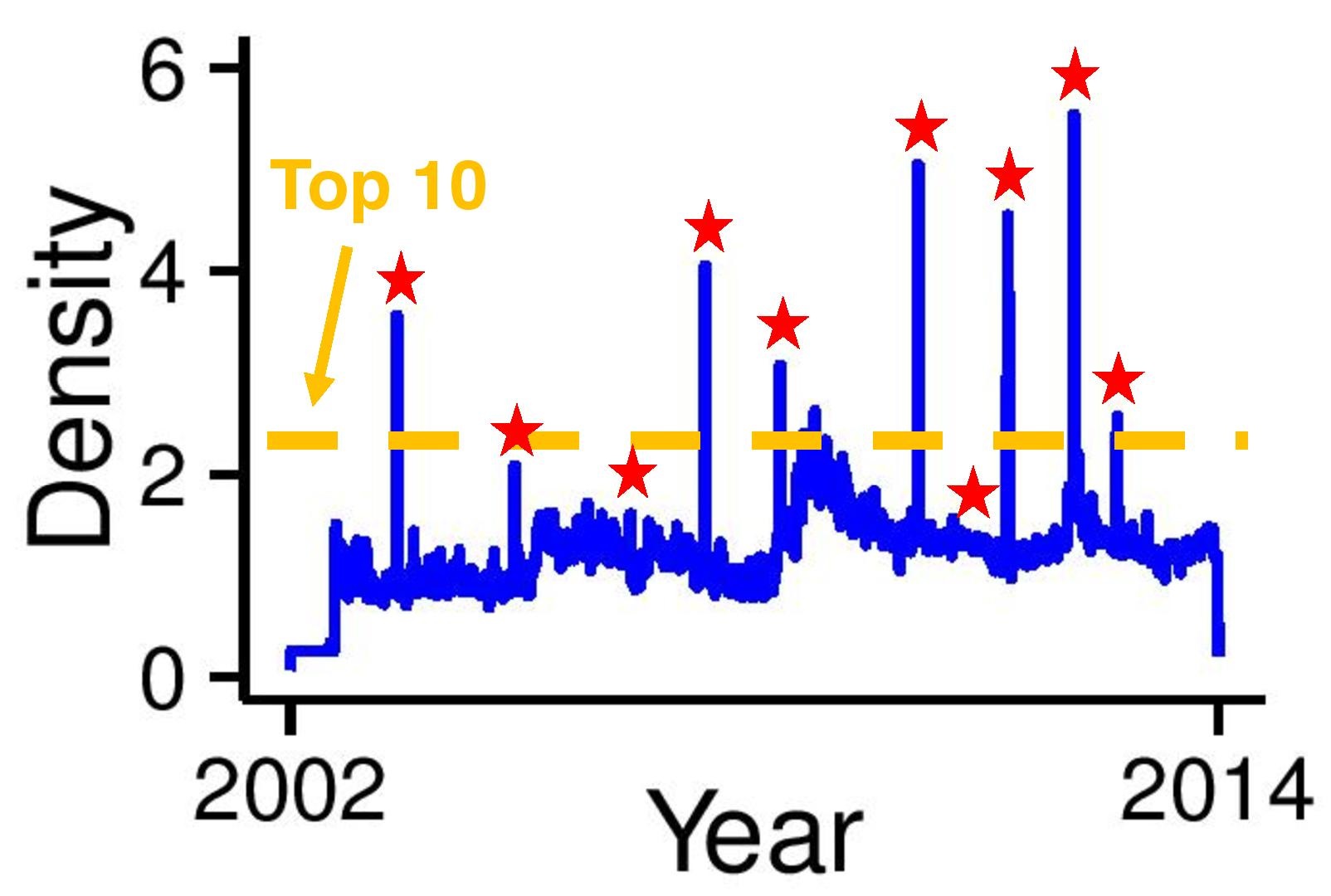}
		\includegraphics[width=0.45\linewidth]{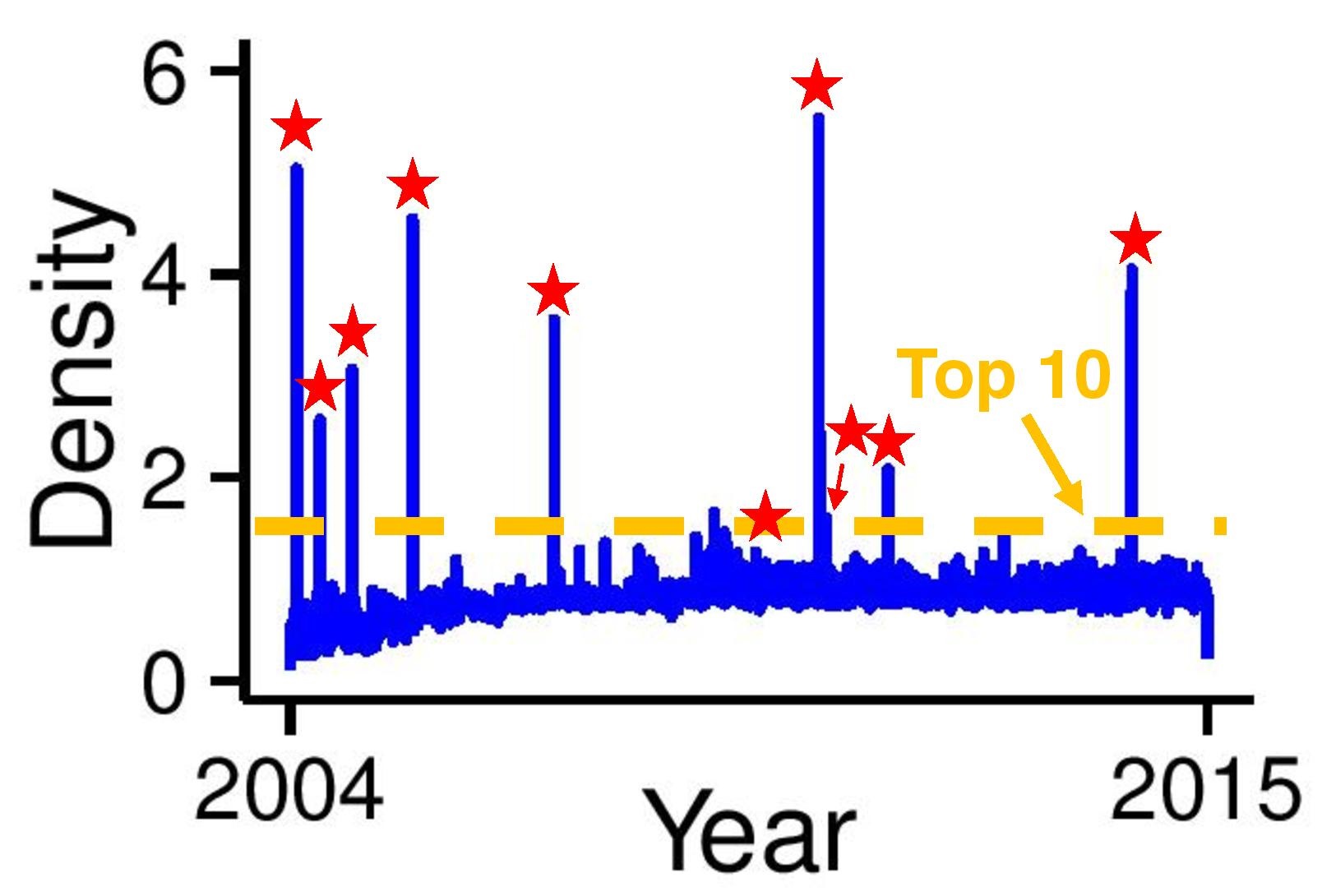}
	} \\
	\vspace{-2mm}
	\subfigure[Comparison with other anomaly detection algorithms]{
		{\small
			\addvbuffer[0pt 4pt]{
				\begin{tabular}{c|c|c}
					\toprule
					& Recall @ Top-10& Recall @ Top-10\\
					& in Yelp & in Android \\
					\midrule
					\methodA & {\bf 0.9} & {\bf 0.7} \\
					\midrule
					Others \cite{hooi2016fraudar,jiang2015general, maruhashi2011multiaspectforensics,shin2016mzoom,shin2017dcube} & 0.0 & 0.0 \\
					\bottomrule
				\end{tabular}
			}
		}
	}
	\caption{\label{fig:review}
		\underline{\smash{\methodA accurately detects small-scale short-}} \underline{\smash{period attacks injected in review datasets.}}
		However, these attacks are overlooked by existing methods.
	}
\end{figure}

\subsubsection{\bf Small-scale Attack Detection in Ratings.} 
\label{sec:exp:effective:rating}
For rating datasets, where ground-truth labels are unavailable,
we assume an attack scenario where fraudsters in a rating site, such as Yelp, utilize multiple user accounts and give the same rating to the same set of items (or businesses) in a short period of time.
The goal of the fraudsters is to boost (or lower) the ratings of the items rapidly.
This lockstep behavior results in a dense subtensor of size $\#fake\_ accounts\times \#target\_items \times 1 \times 1$ in rating datasets whose modes are users, items, timestamps, and ratings.
Here, we assume that fraudsters are not blatant but careful enough to adjust their behavior so that only small-scale dense subtensors are formed.

We injected $10$ such small random dense subtensors of sizes from $3\times3\times1\times1$ to $12\times12\times1\times1$ in Yelp and Android datasets, and compared how many of them are detected by each anomaly-detection algorithm.
As seen in Figure~\ref{fig:review}(a), \methodA (with $\Delta T$=1 time unit in each dataset) clearly revealed the injected subtensors.
Specifically, 9 and 7 among the top $10$ densest subtensors spotted by \methodA indeed indicate the injected attacks in Yelp and Android datasets, respectively.
However, the injected subtensors were not revealed when we simply investigated the
number of ratings in each time unit.
Moreover, as summarized in Figure~\ref{fig:review}(b), none of the injected subtensors was detected\footnote{\small we consider that an injected subtensor is not detected by an algorithm if the subtensor is not included in the top 10 densest subtensors found by the algorithm or it is hidden in a dense subtensor of size at least 10 times larger than the injected subtensor.} by existing algorithms \cite{hooi2016fraudar,jiang2015general, maruhashi2011multiaspectforensics,shin2016mzoom}.
These existing algorithms failed since they either ignore time information \cite{hooi2016fraudar} or only find dense subtensors in the entire tensor \cite{jiang2015general, maruhashi2011multiaspectforensics,shin2016mzoom,shin2017dcube} without using a time window.

\subsubsection{\bf Network Intrusion Detection.} Figure~\ref{fig:crown:tcp} shows the changes of the density of the maintained dense subtensor when we applied \methodA to TCP Dataset with the time window $\Delta T$ = 1 minute.
We found out that the sudden emergence of dense subtensors (i.e., sudden increase in the density) indicates network attacks of various types.
Especially, 
according to the ground-truth labels, all top 15 densest subtensors correspond to actual network attacks.
Classifying each connection as an attack or a normal connection based on the density of the densest subtensor including the connection (i.e., the denser subtensor including a connection is, the more suspicious
the connection is) led to high accuracy with AUC (Area Under the Curve) 0.924.
This was better than MAF (0.514) and comparable with CPD (0.926), \cross (0.923), and \mzoom (0.921).
The result is still noteworthy since \methodA requires only changes in the input tensor within $\Delta T$ time units at a time, while the others require the entire tensor at once.


\subsubsection{\bf Anomaly Detection in Wikipedia.}
\begin{figure*}[t]
	\centering
	\includegraphics[width=0.95\linewidth]{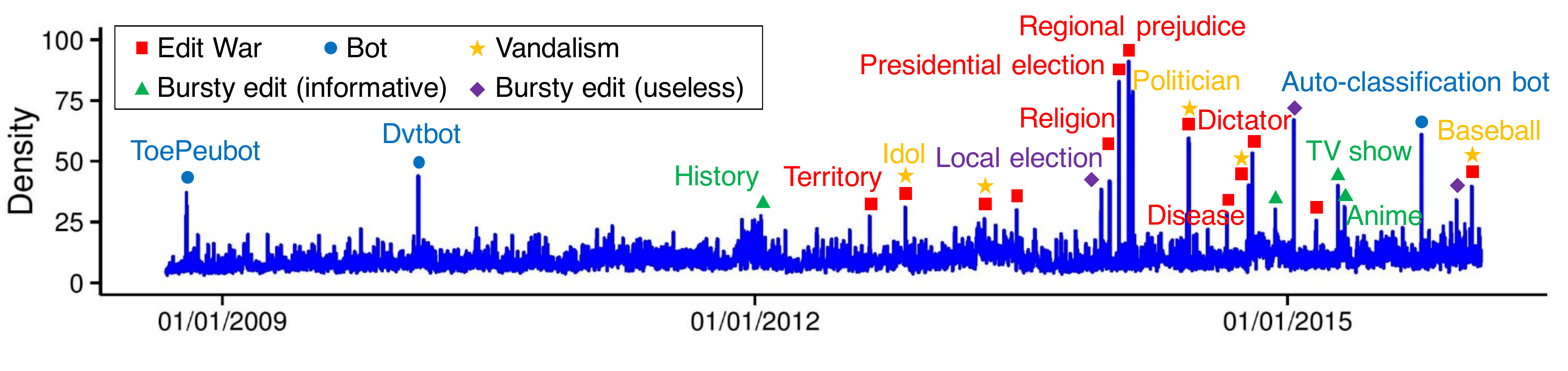} 
	\vspace{-3.5mm}
	\caption{\label{fig:kowiki}
		\underline{\smash{\methodA successfully spots anomalies in Korean Wikipedia.}}
		The sudden appearances of dense subtensors signal actual anomalies of various types including edit wars, bot activities, and vandalism.
		The densest subtensor indicates an edit war where three users edited two pages about regional prejudice 900 times within a day.
	}
\end{figure*}

The sudden appearances of dense subtensors also signal anomalies in Wikipedia edit history.
Figure~\ref{fig:kowiki} depicts the changes of the density of the dense subtensor maintained by \methodA in KoWiki Dataset with the time window $\Delta T$ = 24 hours.
We investigated the detected dense subtensors and found out that most of them corresponded to actual anomalies including edit wars, bot activities, and vandalism.
For example, the densest subtensor, composed by three users and two pages, indicated an edit war where three users edited two pages about regional prejudice 900 times within a day.

%% file: 050related.tex
{\bf Dense Subgraph Detection.} 
For densest-subgraph detection in unweighted graphs,
max-flow-based exact algorithms \cite{goldberg1984finding,khuller2009finding} and greedy approximation algorithms \cite{charikar2000greedy,khuller2009finding} have been proposed. Extensions include adding size bounds \cite{andersen2009finding}, using alternative metrics \cite{tsourakakis2013denser}, finding subgraphs with limited overlap \cite{balalau2015finding,galbrun2016top}, and extending to large-scale graphs  \cite{bahmani2012densest,bahmani2014efficient} and dynamic graphs \cite{epasto2015efficient,mcgregor2015densest,bhattacharya2015space}. 
Other approaches include spectral methods \cite{prakash2010eigenspokes} and frequent itemset mining \cite{seppanen2004dense}. 
Dense-subgraph detection has been widely used to detect fraud or spam in social and review networks \cite{jiang2014catchsync,beutel2013copycatch,shah2014spotting,hooi2016fraudar,shin2016corescope}.

\noindent{\bf Dense Subtensor Detection.} To incorporate additional dimensions and identify lockstep behavior with greater specificity, dense-subtensor detection in multi-aspect data (i.e., tensors) has been considered.
Especially, a likelihood-based approach called \textsc{CrossSpot} \cite{jiang2015general} and a greedy approach giving an accuracy guarantee called \mzoom \cite{shin2016mzoom} were proposed for this purpose. \mzoom was also extended for large datasets stored on a disk or on a distributed file system \cite{shin2017dcube}.
Dense-subtensor detection has been used for network-intrusion detection \cite{maruhashi2011multiaspectforensics,shin2016mzoom,shin2017dcube}, retweet-boosting detection \cite{jiang2015general}, bot detection \cite{shin2016mzoom}, rating-attack detection \cite{shin2017dcube}, genetics applications \cite{saha2010dense}, and formal concept mining \cite{cerf2008data,ignatov2013can}.
However, these existing approaches assume a static tensor rather than a stream of events over time, and do not detect dense subtensors in real time, as they arrive.
We also show their limitations in detecting dense subtensors small but highly concentrated in a short period of time.

\noindent{\bf Tensor Decomposition.} Tensor decomposition such as HOSVD and CPD \cite{kolda2009tensor} can be used to find dense subtensors in tensors, as \textsc{MAF} \cite{maruhashi2011multiaspectforensics} uses CPD for detecting anomalous subgraph patterns in heterogeneous networks. 
Streaming algorithms \cite{sun2006beyond,zhou2016accelerating} also have been developed for CPD and Tucker Decomposition.
However, dense-subtensor detection based on tensor decomposition showed limited accuracy in our experiments (see Section~\ref{sec:exp:accuracy}). 




%% file: 060conclusion.tex
In this work, we propose \methodS, an incremental algorithm for detecting a dense subtensor in a tensor stream, and \methodA, an incremental algorithm for spotting the sudden appearances of dense subtensors.
They have the following advantages:
\begin{compactitem}[$\ \bullet$]
\item {\bf Fast and `any time'}: our algorithms maintain and update a dense subtensor in a tensor stream, which is up to {\em a million times faster} than batch algorithms (Figure~\ref{fig:speed}).
\item {\bf Provably accurate}: 
The densities of subtensors maintained by our algorithms have provable lower bounds (Theorems~\ref{thm:increment:accuracy}, \ref{thm:decrement:accuracy}, \ref{thm:realtime:accuracy}) and are high in practice (Figure~\ref{fig:accuracy}).
\item {\bf Effective}: \methodA successfully detects anomalies, including small-scale attacks, which existing algorithms overlook, in real-world tensors (Figures~\ref{fig:review} and \ref{fig:kowiki}).
\end{compactitem}
{\bf Reproducibility}: The code and data used in the paper are available at \url{http://www.cs.cmu.edu/~kijungs/codes/alert}.

%% file: ACK-kijung.tex
{\small
This material is based upon work
   supported by the National Science Foundation
   under Grant No.
   CNS-1314632 
   and IIS-1408924. 
   Research was sponsored by the Army Research Laboratory 
   and was accomplished under Cooperative Agreement Number W911NF-09-2-0053. 
   Kijung Shin is supported by KFAS Scholarship. Jisu Kim is supported by Samsung Scholarship.
   Any opinions, findings, and conclusions or recommendations expressed in this
   material are those of the author(s) and do not necessarily reflect the views
   of the National Science Foundation, or other funding parties.
   The U.S. Government is authorized to reproduce and 
   distribute reprints for Government purposes notwithstanding 
   any copyright notation here on.}